\documentclass{osa-article}

\usepackage{graphicx}% Include figure files
\usepackage{dcolumn}% Align table columns on decimal point
\usepackage{bm}% bold math
\usepackage{subfigure}
\usepackage{float}
\usepackage{color}
\usepackage{mathrsfs}
\usepackage{amstext}
\usepackage{booktabs}
\usepackage{siunitx}
\usepackage{overpic}
\usepackage{threeparttable}

\journal{osajournal}
\setprjcopyright

\articletype{Research Article}
\begin{document}

\title{Ultrasensitive Measurement of Angular Rotations via Hermite-Gaussian Pointer}

\author{Binke Xia, Jingzheng Huang\authormark{*}, Hongjing Li, Miaomiao Liu, Tailong Xiao, Chen Fang and Guihua Zeng\authormark{$\dagger$}}

\address{State Key Laboratory of Advanced Optical Communication Systems and Networks, Institute for Quantum Sensing and Information Processing, Shanghai Jiao Tong University, Shanghai 200240, China}

\email{\authormark{*}jzhuang1983@sjtu.edu.cn, \authormark{$\dagger$}ghzeng@sjtu.edu.cn}



\begin{abstract}
Exploring high sensitivity on the measurement of angular rotations is an outstanding challenge in optics and metrology. In this work, we employ the mn-order Hermite-Gaussian beam in the weak measurement scheme with an angular rotation interaction, where the rotation information is taken by another HG mode state completely after the post-selection. By taking a projective measurement on the final light beam, the precision of angular rotation is improved by a factor of 2mn+m+n. For verification, we perform an optical experiment where the minimum detectable angular rotation improves $\sqrt{15}$-fold with HG55 mode over that of HG11 mode, and achieves a sub-$\si{\micro rad}$ scale of the measurement precision. Our theoretical framework and experimental results not only provide a more practical and convenient scheme for ultrasensitive measurement of angular rotations, but also contribute to a wide range of applications in quantum metrology.
\end{abstract}

%%%%%%%%%%%%%%%%%%%%%%%%%%  body  %%%%%%%%%%%%%%%%%%%%%%%%%%
\section{Introduction}
Measuring the angular rotations with high sensitivity has an increasing of interest recently, for its growing potential in a wide range of optical science and applications. For example, precise measurement of rotations plays a vital role in atom interferometer gyroscopes\cite{PhysRevLett.107.133001}, optical tweezers\cite{Padgett_2011}, rotational Doppler effect\cite{PhysRevLett.81.4828,doi:10.1126/science.1239936,Zhang:20} and magnetic field measurements\cite{doi:10.1063/1.4923446}. Traditionally, the basic laser beam with Gaussian profile is incapable to take angular rotations because it is rotational symmetry\cite{Pampaloni_2004}. Motivating by the studies of light endowed with orbital angular momentum (OAM)\cite{PhysRevA.45.8185}, some related efforts are proposed to increase the sensitivity of angular rotations measurement. But it is worth to note that the pure OAM lights like Laguerre-Gaussian (LG) beams are still rotational symmetry, therefore quantum resources are involved for angular rotations measurement in addition, such as quantum entanglement of high OAM values\cite{doi:10.1126/science.1227193} and N00N states in the OAM bases\cite{PhysRevA.83.053829,Bouchard:17}. Recently, Vincenzo $\textit{et al.}$ also proposed a scheme on the rotation measurements with a $\SI{}{\micro rad}$-scaling sensitivity by utilizing the classical entangled formalism of OAM and polarization\cite{Ambrosio_2013}. However, these schemes are complicated to implement, for example, quantum resources are usually difficult to generate\cite{Afek879,Xiang_2011} and fragile in noise\cite{Higgins_2007}, and a customized $q$-plate is necessary for generating OAM-polarization entangled formalism\cite{Ambrosio_2013,Barboza_2022}. To explore the more practical and simple protocol for precise measurement of angular rotations, Omar $\textit{et al.}$ have reported a weak value amplification scheme with experimental precision of  $\ang{0.4}(\approx\SI{7}{\milli rad})$, where the light beam with angular Gaussian profile is employed\cite{PhysRevLett.112.200401}.

Those previous works concentrated on the precision improvement via higher OAM values, but we find that the ultimate precision on rotation measurement is decided by the variance of OAM distribution instead of OAM value. Therefore, we employ the Hermite-Gaussian (HG) pointer to achieve an ultra-high sensitivity on the measurement of angular rotations because of its large OAM variances. Previously, the $n$-order HG pointer was employed on the measurement of spatial displacement\cite{PhysRevA.74.053823,doi:10.1063/1.4869819}, where the corresponding quantum Cram$\acute{\mathrm{e}}$r-Rao (QCR) bound\cite{Holevo2011} was the improved linearly with mode number $n$. In this work, we employ the $mn$-order Hermite-Gaussian pointer in a rotational-coupling weak measurement scheme. After the post-selection, the information of angular rotation is taken by an HG mode state which is orthogonal to the initial pointer state. Then projecting the final light beam to the state taken angular rotations, the quantum limit precision of rotation measurement can be achieved, which is enhanced with a factor of $2mn+m+n$.

For demonstration, we set up an optical experiment to implement the precision measurement of angular rotation. Instead of the tomography of OAM distribution in \cite{PhysRevLett.112.200401}, we demodulate the angular rotations from the projection intensity directly, and the imaginary weak value is also not necessary. Especially, the measurement precision in our experiment achieves $\SI{0.89}{\micro rad}$ with the 5$\times$5-order HG beam. Our results shed new light on the precise measurement of angular rotation, and have potential for optical metrology, remote sensing, biological imaging, and navigation systems\cite{doi:10.1126/science.1239936,Rosales_2013,PhysRevLett.110.043601}.

\section{Theoretical model}
\subsection{Enhanced quantum limit via HG pointer}
\begin{figure}[h]
	\centering
	\includegraphics[scale=0.5]{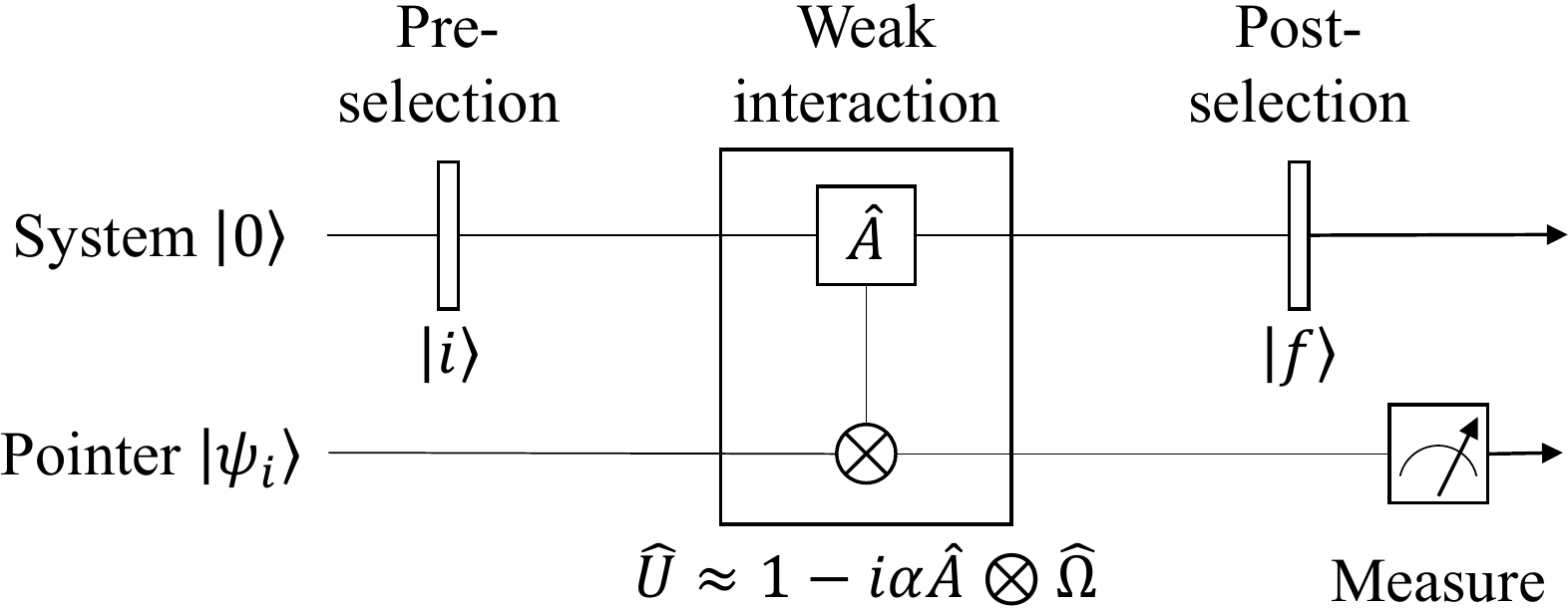}
	\caption{\label{fig:1} Post-selected weak measurement scheme.}
\end{figure}

To be clear, we first consider the general weak measurement process with post-selection, as is depicted in Fig. \ref{fig:1}. For simplicity, we consider a two level system with initial state $|i\rangle$ and a pointer with initial state $|\psi_{i}\rangle$. They couple together during the weak interaction procedure with an impulse Hamiltonian $\hat{H}_{I}=\delta\left(t-t_{0}\right)\alpha\hat{A}\otimes\hat{\Omega}$, where $\hat{\Omega}$ is a translation operator on pointer, and $\alpha$ is the corresponding interaction strength. Here $\hat{A}$ is a Pauli operator on the two-level system. In weak measurement scheme, the interaction strength $\alpha\ll 1$, then the unitary evolution operator of weak interaction procedure can be approximately calculated as $\hat{U}=\exp\left(-\mathrm{i}\int\hat{H}_{I}\mathrm{d}t\right)\approx 1-\mathrm{i}\alpha\hat{A}\otimes\hat{\Omega}$. (Without loss of generality, we adopt units making $\hbar=1$ in this article.)

To individually read out the measurement information from the pointer, we post-select the system by state $|f\rangle$, turn the final state in whole to be $|\psi_{f}\rangle|f\rangle$, where $|\psi_{f}\rangle\approx\mathcal{N}\left(1-\mathrm{i}\alpha A_{w}\hat{\Omega}\right)|\psi_{i}\rangle$ is the pointer's final state, and $\mathcal{N} = 1/\sqrt{1+2\alpha\mathrm{Im}\left(A_{w}\right)\langle\hat{\Omega}\rangle_{i}+\alpha^{2}\left|A_{w}\right|^{2}\langle\hat{\Omega}^{2}\rangle_{i}}$ is the normalized factor with $\langle\hat{\Omega}\rangle_{i}=\langle\psi_{i}|\hat{\Omega}|\psi_{i}\rangle$, $\langle\hat{\Omega}^{2}\rangle_{i}=\langle\psi_{i}|\hat{\Omega}^{2}|\psi_{i}\rangle$. And $A_{w}$ is weak value calculated by $A_{w}=\langle f|\hat{A}|i\rangle/\langle f|i\rangle$\cite{PhysRevLett.60.1351,PhysRevA.76.044103}.

To analyze the estimating precision in our weak measurement scenario, we employ the quantum Fisher information (QFI) as a figure of merit\cite{Matsumoto_2002,DEMKOWICZDOBRZANSKI2015345}. The QFI of final pointer state $|\psi_{f}\rangle$ about interaction strength parameter $\alpha$ can be calculated as $\mathcal{Q}\left(\alpha\right)\approx 4\left|A_{w}\right|^{2}\langle\Delta\hat{\Omega}^{2}\rangle_{i}$, where $\langle\Delta\hat{\Omega}^{2}\rangle_{i}=\langle\hat{\Omega}^{2}\rangle_{i}-\langle\hat{\Omega}\rangle_{i}^{2}$. For $N$ classical measured samples, the variance of estimator $\hat{\alpha}$ satisfies the QCR inequality $\delta\hat{\alpha}^{2}\ge 1/N\mathcal{Q}\left(\alpha\right)$, which leads to an uncertainty relation:
\begin{equation}
	\delta\hat{\alpha}^{2}\langle\Delta\hat{\Omega}^{2}\rangle_{i} \ge \frac{1}{4N\left|A_{w}\right|^{2}} \label{eq:1}
\end{equation}

In this work, we apply the weak measurement scheme to the measurement of angular rotation. Thus, the interaction strength corresponds to the rotation angle of pointer, and the translation operator $\hat{\Omega}=\hat{L}_{z}$ is the orbital angular momentum operator. Traditionally, the laser beam with Gaussian profile is widely used in weak measurement, the corresponding spatial wave function is $\psi_{\mathrm{G}}(x,y)=\frac{1}{\sqrt{2\pi\sigma_{0}^{2}}}\exp\left(-\frac{x^{2}+y^{2}}{4\sigma_{0}^{2}}\right)$, where $\sigma_{0}^{2}$ is the spatial variance of Gaussian beam. Obviously, $\langle\Delta\hat{L}_{z}^{2}\rangle_{\mathrm{G}}=0$ because of the rotational symmetry of Gaussian beam. Thus, it is necessary to devise appropriate pointer for rotation measurement. In addition, revealing from Eq. \ref{eq:1}, increasing the variance of OAM of the pointer is benefit for higher precision on measuring angular rotations. 

For this reason, we employ the $mn$-order HG beam as initial pointer for the measurement of angular rotations, where $m$ and $n$ are the transverse mode numbers of $x$-component and $y$-component separately. Though the $mn$-order HG beam takes zero-mean OAM\cite{Pampaloni_2004}, its variance of OAM distribution increases quadratically with the mode numbers:
\begin{equation}
	\langle\Delta\hat{L}_{z}^{2}\rangle_{mn} = 2mn+m+n \label{eq:2}
\end{equation}
which is going to be derived following. Then the quantum limit of rotation measurement with $mn$-order HG beam is given as:
\begin{equation}
	\delta\hat{\alpha}^{2} \ge \frac{1}{4N\left|A_{w}\right|^{2}(2mn+m+n)} \label{eq:3}
\end{equation}
which is derived from the QCR inequality in Eq. \ref{eq:1}, and the ultimate precision is improved quadratically with mode numbers $m$ and $n$. Besides the improvement on the quantum limit of rotation measurement, employing HG beams also provides more convenient way to implement the optimal measurement for demodulating angular rotations in a practical system.

\subsection{Operator algebra of HG beams}
In detail, we can relate the HG beams to the harmonic oscillators (HO) here. The wave function of HG beam is\cite{247715}:
\begin{equation}
	u_{mn}\left(x,y,z\right) = \exp\left[\frac{\mathrm{i}k(x^{2}+y^{2})}{2q(z)}-\mathrm{i}(m+n+1)\chi(z)\right]\frac{\sigma_{0}}{\sigma(z)}\psi_{mn}\left[\frac{\sigma_{0}}{\sigma(z)}x,\frac{\sigma_{0}}{\sigma(z)}y\right] \label{eq:4}
\end{equation}
where the three z-dependent parameters, spatial variance $\sigma^{2}$, Gouy phase $\chi$ and radius of curvature of
the wavefront $q$ can be determined by equalities
\begin{equation}
	\frac{1}{2\sigma^{2}(z)}-\frac{\mathrm{i}k}{q(z)} = \frac{k}{b+\mathrm{i}z},\quad\tan\chi(z)=\frac{z}{b} \label{eq:5}
\end{equation}
where $k$ is the wave number and $b$ is the Rayleigh range\cite{doi:10.1098/rsta.2015.0443}. And $\psi_{mn}\left(x,y\right)$ in Eq. \ref{eq:4} is the 2-D harmonic Hermite-Gaussian function
\begin{equation}
	\psi_{mn}(x,y) = \frac{H_{m}(\frac{x}{\sqrt{2}\sigma_{0}})H_{n}(\frac{y}{\sqrt{2}\sigma_{0}})}{\sqrt{2^{m+n+1}\pi\sigma_{0}^{2}m!n!}}\exp\left(-\frac{x^{2}+y^{2}}{4\sigma_{0}^{2}}\right) \label{eq:6}
\end{equation}
where $H_{n}$ is the n-order Hermite polynomial.

From the view of quantum mechanics, wave function $\psi_{mn}\left(x,y\right)$ is the time-independent solution for Schr$\ddot{\mathrm{o}}$dinger equation of 2-D harmonic oscillators:
\begin{equation}
	\mathrm{i}\frac{\partial\psi}{\partial t} = \left[\sigma_{0}^{2}\left(\hat{p}_{x}^{2}+\hat{p}_{y}^{2}\right)+\frac{1}{4\sigma_{0}^{2}}\left(\hat{x}^{2}+\hat{y}^{2}\right)\right]\psi \label{eq:7}
\end{equation}
For eigenvalue $E_{mn}=\left(m+n+1\right)$, the corresponding eigenket can be obtained as:
\begin{equation}
	|m,n\rangle = \iint\mathrm{d}x\,\mathrm{d}y\,\psi_{mn}(x,y)|x,y\rangle \label{eq:8}
\end{equation}

Here, we denote the $mn$-order HG beam state as
\begin{equation}
	|u_{mn}(z)\rangle = \iint\mathrm{d}x\,\mathrm{d}y\,u_{mn}(x,y)|x,y\rangle \label{eq:9}
\end{equation}
Obviously, $|u_{mn}(0)\rangle=|m,n\rangle$. Defining the creation (annihilation) operators for the HO state $|m,n\rangle$:
\begin{align}
	\hat{a}_{x}^{\dagger}|m,n\rangle = \sqrt{m+1}|m+1,n\rangle,\;&\hat{a}_{x}|m,n\rangle = \sqrt{m}|m-1,n\rangle \label{eq:10}\\
	\hat{a}_{y}^{\dagger}|m,n\rangle = \sqrt{n+1}|m,n+1\rangle,\;&\hat{a}_{y}|m,n\rangle = \sqrt{n}|m,n-1\rangle \label{eq:11}
\end{align}

HG beam states are $z$-dependent, and its wave functions are the solutions of the paraxial wave equation:
\begin{equation}
	-2\mathrm{i}k\frac{\partial}{\partial z}u(x,y,z) = \left(\frac{\partial^{2}}{\partial x^{2}}+\frac{\partial^{2}}{\partial y^{2}}\right)u(x,y,z) \label{eq:12}
\end{equation}
which can be rewritten as:
\begin{equation}
	\frac{\mathrm{d}}{\mathrm{d}z}|u_{n}(z)\rangle = -\frac{\mathrm{i}}{2k}\left(\hat{p}_{x}^{2}+\hat{p}_{y}^{2}\right)|u_{n}(z)\rangle \label{eq:13}
\end{equation}
This equation has the formal solution $|u_{mn}(z)\rangle=\hat{U}(z)|u_{mn}(0)\rangle$ with the propagation operator
\begin{equation}
	\hat{U}(z) = \exp\left[-\frac{\mathrm{i}}{2k}\left(\hat{p}_{x}^{2}+\hat{p}_{y}^{2}\right)z\right] \label{eq:14}
\end{equation}
Thus, the $z$-dependent mode creation (annihilation) operators can be derived by:
\begin{align}
	\hat{a}_{x}(z) &= \hat{U}(z)\hat{a}_{x}\hat{U}^{\dagger}(z),\;\hat{a}_{x}^{\dagger}(z) = \hat{U}(z)\hat{a}_{x}^{\dagger}\hat{U}^{\dagger}(z) \label{eq:15} \\
	\hat{a}_{y}(z) &= \hat{U}(z)\hat{a}_{y}\hat{U}^{\dagger}(z),\;\hat{a}_{y}^{\dagger}(z) = \hat{U}(z)\hat{a}_{y}^{\dagger}\hat{U}^{\dagger}(z) \label{eq:16}
\end{align}
Hence, the momentum and position operators can be obtained by these $z$-dependent creation and annihilation operators:
\begin{equation}
	\label{eq:17}
	\left\{
	\begin{split}
		&\hat{p}_{x} = -\frac{\mathrm{i}}{2\sigma_{0}}\left[\hat{a}_{x}(z)-\hat{a}_{x}^{\dagger}(z)\right] \\
		&\hat{x} = \sigma_{0}\left[\hat{a}_{x}(z)+\hat{a}_{x}^{\dagger}(z)\right]+\frac{z\sigma_{0}}{\mathrm{i}b}\left[\hat{a}_{x}(z)-\hat{a}_{x}^{\dagger}(z)\right] \\
		&\hat{p}_{y} = -\frac{\mathrm{i}}{2\sigma_{0}}\left[\hat{a}_{y}(z)-\hat{a}_{y}^{\dagger}(z)\right] \\
		&\hat{y} = \sigma_{0}\left[\hat{a}_{y}(z)+\hat{a}_{y}^{\dagger}(z)\right]+\frac{z\sigma_{0}}{\mathrm{i}b}\left[\hat{a}_{y}(z)-\hat{a}_{y}^{\dagger}(z)\right]
	\end{split}
	\right.
\end{equation}

Moreover, it is easy to determine that the momentum operators $\hat{p}_{x} = -\frac{\mathrm{i}}{2\sigma_{0}}\left(\hat{a}_{x}-\hat{a}_{x}^{\dagger}\right)$ and $\hat{p}_{y} = -\frac{\mathrm{i}}{2\sigma_{0}}\left(\hat{a}_{y}-\hat{a}_{y}^{\dagger}\right)$ for 2-D HO states, which have the same expression with that of HG beam states. In another word, the result of inflicting displacement on the $mn$-order HG beam state is same as that of $mn$-order HO state. Then we can derive the OAM operator $\hat{L}_{z}$ by Eq. \ref{eq:17}:
\begin{equation}
	\hat{L}_{z} = \hat{x}\hat{p}_{y}-\hat{y}\hat{p}_{x} =  \mathrm{i}\left[\hat{a}_{x}(z)\hat{a}_{y}^{\dagger}(z)-\hat{a}_{x}^{\dagger}(z)\hat{a}_{y}(z)\right] \label{eq:18}
\end{equation}
Obviously, the OAM variance of $mn$-order HG beam $\langle\Delta\hat{L}_{z}^{2}\rangle_{mn}=2mn+m+n$ is $z$-independent, and the OAM operator $\hat{L}_{z}$ is also $z$-independent because $\hat{L}_{z}\equiv\hat{U}(z)\hat{L}_{z}\hat{U}^{\dagger}(z)$. Thus, the $mn$-order HG beam state is equivalent to the $mn$-order HO state in the scenario of rotation measurement.

\subsection{Saturating quantum limit via projective measurement}
Taking the initial pointer state as $|\psi_{i}\rangle=|m,n\rangle$, then the final pointer state can be calculated as:
\begin{equation}
	|\psi_{f}\rangle \approx |m,n\rangle+A_{w}\alpha\left[\sqrt{m\left(n+1\right)}|m-1,n+1\rangle-\sqrt{\left(m+1\right)n}|m+1,n-1\rangle\right] \label{eq:19}
\end{equation}
where the rotation parameter are carried by a HG mode state:
\begin{equation}
	|\psi_{\hat{L}}\rangle = \frac{1}{\sqrt{2mn+m+n}}\left[\sqrt{m\left(n+1\right)}|m-1,n+1\rangle-\sqrt{\left(m+1\right)n}|m+1,n-1\rangle\right] \label{eq:20}
\end{equation}
which is a superposition state of pointer's adjacent modes $|m-1,n+1\rangle$ and $|m+1,n-1\rangle$.

In a complete metrological process, a classical measurement strategy is necessary for the final state to read out the unknown parameters\cite{Liu_2019}. In this case, the final estimating precision of angular rotation is evaluated by the classical Fisher information (CFI):
\begin{equation}
	\mathcal{F}(\alpha) = \sum_{\lambda}\frac{1}{\langle\psi_{f}|\hat{\Pi}_{\lambda}|\psi_{f}\rangle}\left(\frac{\partial}{\partial\alpha}\langle\psi_{f}|\hat{\Pi}_{\lambda}|\psi_{f}\rangle\right)^{2} \label{eq:21}
\end{equation}
where $\hat{\Pi}=\left\{\hat{\Pi}_{\lambda}\,\big|\,\hat{\Pi}\ge 0,\sum_{\lambda}\hat{\Pi}_{\lambda}=\hat{\mathbb{I}}\right\}$ is a set of positive-operator-valued measure (POVM). Then the practical precision of angular rotation is limited by the classical Cram$\acute{\mathrm{e}}$r-Rao (CCR) bound $\delta\hat{\alpha}^{2}\ge 1/N\mathcal{F}(\alpha)$.
\begin{figure}[h]
	\centering
	\includegraphics[width=0.8\linewidth]{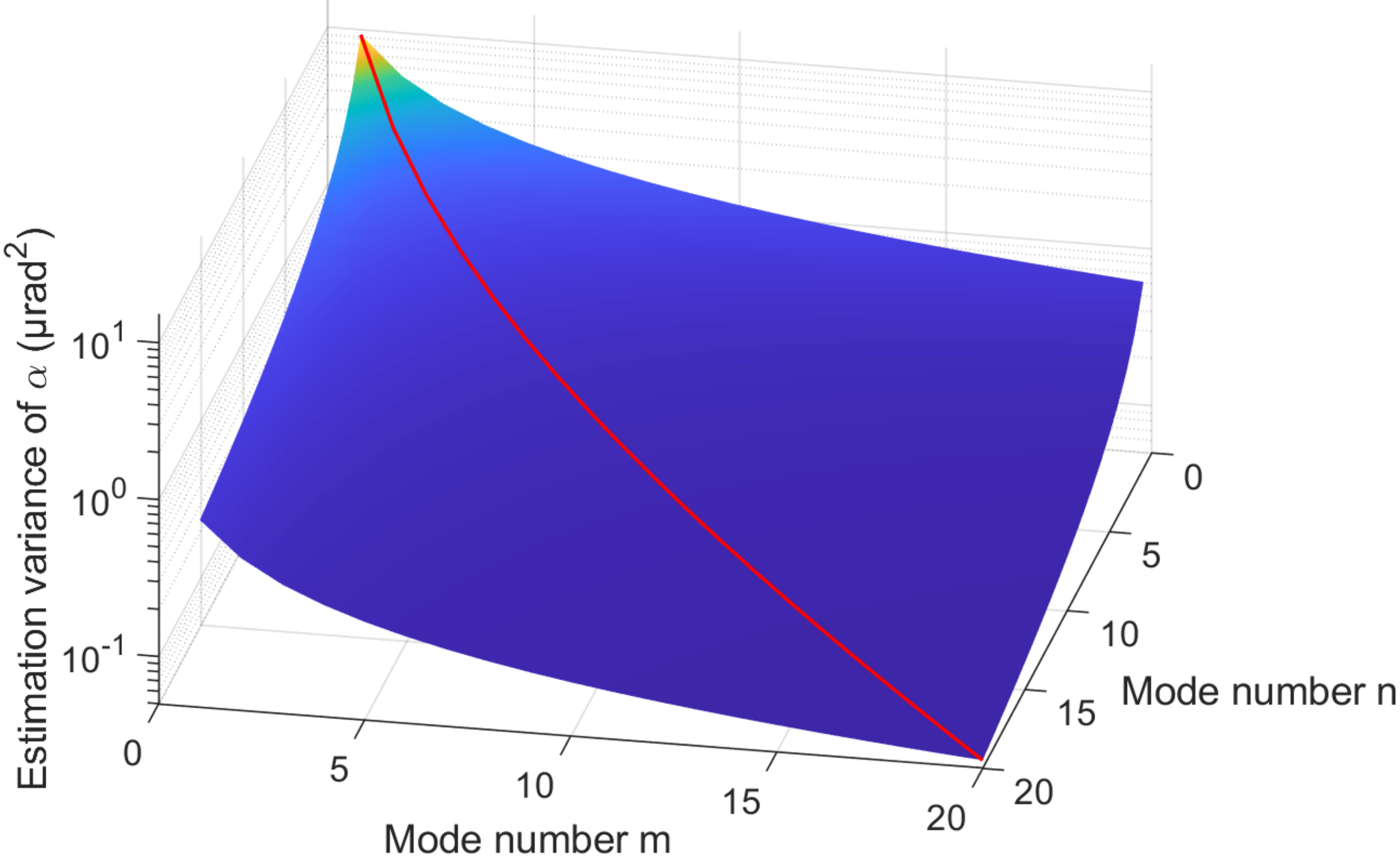}
	\caption{\label{fig:2} Lower bound of estimation variances $\delta\hat{\alpha}^{2}$ with different mode numbers under projective measurement method. The measured photons number is set as $N=\num{4.04d7}$, and the weak value is set as $A_{w}=\cot\ang{5}\approx 11$. The x-axis and y-axis are the mode number of $m$ and $n$ separately, and z-axis is the estimation variance of parameter $\alpha$. The red line in this figure is the CCR bound at $m=n$, where the precision limit is improved fastest.}
\end{figure}
Basically, the CCR bound of single parameter is capable to saturate the quantum limit given by QCR inequality via devising an optimal measurement strategy\cite{Liu_2019}. For the measurement of angular rotation, the tomography of OAM distributions $\hat{\Pi}_{\mathrm{OAM}}=\left\{|l\rangle\langle l|\,\big|\,l\in\mathbb{Z}\right\}$ was usually chosen as the optimal POVM traditionally\cite{PhysRevLett.112.200401}, where $|l\rangle$ is the eigenket of OAM operator. However, the complete tomography requires infinite projective measurements on different eigenkets $|l\rangle$ for final pointer theoretically. In our scheme with $mn$-order HG pointer, a single projection for final pointer on the state $|\psi_{\hat{L}}\rangle$ is capable to demodulating the angular rotations, tomography of OAM spectrum or HG mode spectrum is no more required. Especially, no matter how large the mode number of initial HG pointer is, the optimal POVM on the final pointer is a single projective measurement $\hat{\Pi}_{\mathrm{HG}}=\left\{\hat{\Pi}_{\hat{L}}=|\psi_{\hat{L}}\rangle\langle\psi_{\hat{L}}|,\,\hat{\mathbb{I}}-\hat{\Pi}_{L}\right\}$. Then the CFI of rotation parameter can be calculated as $\mathcal{F}(\alpha)=4\left|A_{w}\right|^{2}(2mn+m+n)$, which leads to the CCR bound $\delta\hat{\alpha}^{2}\ge 1/4N\left|A_{w}\right|^{2}(2mn+m+n)$ saturating the corresponding QCR bound in Eq. \ref{eq:3}. To visualize the dependency of the theoretical precision limit on the mode numbers $m$ and $n$, we illustrate the CCR bound of parameter $\alpha$ with different mode numbers under projective measurement method in Fig. \ref{fig:2}, where the measured photons number is set as $N=\num{4.04d7}$, and the weak value is set as $A_{w}=\cot\ang{5}\approx 11$. Besides, we also plot the CCR bound at $m=n$ with the red line in Fig. \ref{fig:2}, where the precision limit is improved fastest, and the dependency of the sensitivity on the factor of $2mn+m+n$ is evident.

\section{Experimental scheme}
\subsection{Experimental materials and setup}
\begin{figure*}[htb]
	\centering
	\includegraphics[width=\linewidth]{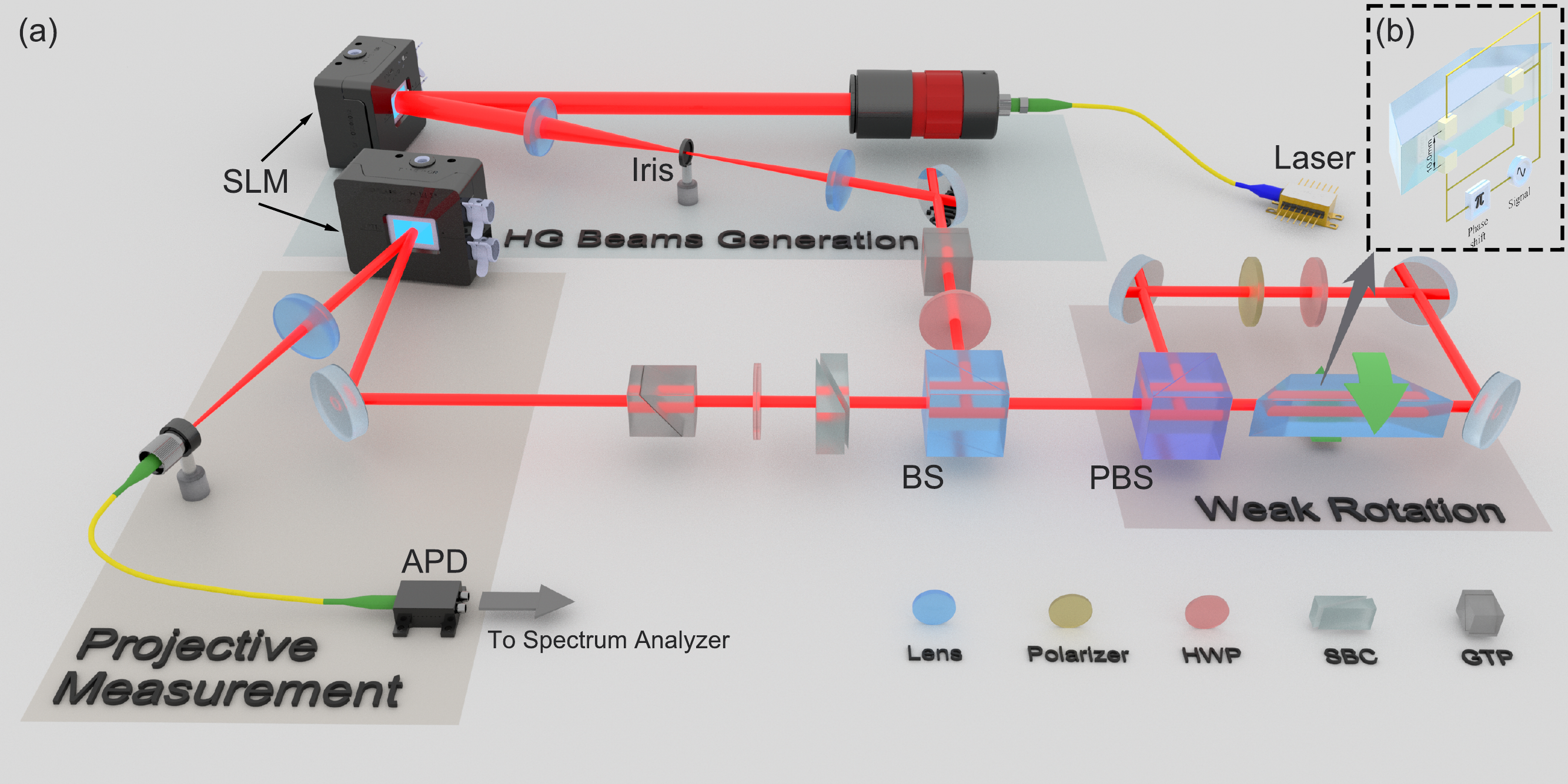}
	\caption{\label{fig:3} Diagram of experimental setup. (a) The $mn$-order HG beam is converted from a expanded Gaussian beam of $\SI{780}{\nm}$ laser by a SLM and a spatial filter system. The pre-selection is implemented by a Glan-Taylor polarizer (GTP) and a half-wave plate (HWP). And a polarized Sagnac interferometer is employed to implement the weak interaction procedure, where the inverse rotation signals are introduced by a Dove prism. Then a Soleil-Babinet compensator (SBC), a HWP and a GTP are used to implement the post-selection. Finally, another SLM with a Fourier transfer lens are used to implement the projective measurement, where the successful projected photons are collected by an APD with a SMF. (b) Dove prism with PZT chips and generation method of rotation signal. There are 4 PZT chips pasted on the reflection side of prism with a $2\times 2$ array, where the vertical distance of the PZT array is $\SI{10}{\mm}$.}
\end{figure*}
To experimentally verify that the enhancement on rotation measurement with HG pointer, we setup a practical optical system to implement it, as is shown in Fig. \ref{fig:3}. A light beam from the laser working at $\SI{780}{\nm}$ is expanded and then converted to $mn$-order Hermite-Gaussian mode via a spatial light modulator (SLM) and a spatial filter system\cite{Clark:16}. And here beam's polarization states $|H\rangle$ and $|V\rangle$ are set as the basis of the two-level system. Here we employ a Dove prism to introduce a pair of inverse weak rotations $\alpha$ for $|H\rangle$ and $|V\rangle$ component in a polarizing Sagnac interferometer. The Pauli operator is denoted as $\hat{A}=|H\rangle\langle H|-|V\rangle\langle V|$. In the post-selected weak measurement scheme, pre-selection and post-selection states are nearly orthogonal to amplify the estimated parameter\cite{Hosten787,PhysRevLett.102.173601,PhysRevLett.112.200401,Hallaji_2017,doi:10.1063/1.5027117,PhysRevA.97.063818}. Thus, we choose $|i\rangle=\frac{1}{\sqrt{2}}\left(|H\rangle+|V\rangle\right)$ and $|f\rangle=\cos\left(\frac{\pi}{4}-\varepsilon\right)|H\rangle-\sin\left(\frac{\pi}{4}-\varepsilon\right)|V\rangle$, where $\varepsilon\ll 1$. Thus,  weak value $A_{w}=\cot\varepsilon$. Considering $N$ measurement samples (effective measured photons number in experiment), the minimum detectable rotation $\alpha$ given by QCR bound is:
\begin{equation}
	\alpha_{\min}^{\mathrm{QCR}} = \frac{1}{\sqrt{2mn+m+n}} \frac{1}{2|\cot\varepsilon|\sqrt{N}} \label{eq:22}
\end{equation}
which is significantly improved by the spatial mode numbers of HG beams.

The laser employed in this experiment is a Distributed Bragg Reflector (DBR) Single-Frequency Laser of of Thorlabs Inc. (part number: DBR780PN), which works at $\SI{780}{\nm}$ with $\SI{1}{\MHz}$ typical linewidth. To generate the high-order HG beams, we used a SLM of Hamamatsu Photonics (part number: X13138-02), which has $1272\times 1024$ pixels with $\SI{12.5}{\um}$ pixel pitch. The focal length of the Fourier lens in the 4-f system is $\SI{5}{\cm}$. A $\SI{200}{\um}$ square pinhole is used as the spatial filter.

In this work, we set up a polarized Sagnac interferometer to introduce a pair of inverse rotation signals for horizontal and vertical polarization states. However, the extinction ratio of the reflection port of the polarizing beamsplitter (PBS) cube (part number CCM1-PBS25-780/M of Thorlabs Inc.) is from 20:1 to 100:1 in practice, which deteriorates the degree of polarization of the output beam. Hence, we added a polarizer behind the reflection port of the PBS to improve the degree of polarization.  And a HWP, whose optic axial is at $\ang{45}$ angle to the horizontal plane, was employed to exchange the polarization states in the clockwise loop and counterclockwise loop.

In our experimental scheme, the Dove prism is driven by piezoelectric transducer (PZT) chips, and we exert an $f=\SI{1}{\kHz}$ cosine driving signal on the PZT to generate the tiny rotation signal. Here, we pasted 4 PZT chips on the reflection side of Dove prism, as is illustrated in Fig. \ref{fig:3}(b). The 4 PZT chips arrange as a $2\times 2$ array, and the vertical distance of this array is $\SI{10}{\mm}$. Here, we used the NAC2013 PZT chip of Core Tomorrow Company, which shifts $\SI{22}{\nm}$ with 1V driving voltage. We exert cosine signals (with 1/2 amplitude DC bias) on the PZT chips, where a $\pi$-phase difference in introduced between the top-row PZT chips and bottom-row PZT chips. Therefore, a cosine driving signal with $\SI{1}{\V}$ peak-to-peak voltage corresponds to a $\SI{2.2}{\micro rad}$ maximum rotation of Dove prism, which leads to a $\SI{4.4}{\micro rad}$ transverse rotation of input light beam. Besides this cosine driving signal, the initial rotation bias of Dove prism, which is denoted as $\alpha_{0}$ and on the $\si{\milli rad}$ scale, is non-negligible. Thus, the total rotation is $\alpha_{tot}=\alpha_{0}+\alpha\cos(2\pi ft)$, and it is easy to determine that $\alpha\ll\alpha_{0}\ll 1$.

After the post selection, another SLM is employed to project the final pointer to carrying state $|\psi_{\hat{L}}\rangle$ with a Fourier transfer lens and a spatial filtering from single mode fiber (SMF) coupling detected photons to an avalanche photodiode (APD, part number: APD440A of Thorlabs Inc.), which has maximum conversion gain of $\SI{2.65d9}{\V/\W}$ and $\SI{100}{\kHz}$ bandwidth. Then the detected voltage signal was analyzed by the spectrum analyzer module of Moku:Lab, which is a reconfigurable hardware platform produced by Liquid instruments. The resolution bandwidth (RWB) of spectrum analyzer was $\SI{9.168}{\Hz}$ in our experiment, which leads to the detecting time of $\tau=\SI{109.08}{\ms}$.

\subsection{Experimental results}
In practice, before exerting the driving signal, we project the final pointer to state $|m,n\rangle$ to fix the measured photons number $N$ for different HG pointers. In the experiment, the detected power of APD is fixed as $I_{0}=\SI{94.34}{\pico\W}$ at the this beforehand projection step. Theoretically, the detected optical power is given as $I_{0}=\gamma N/\tau$, where $\gamma$ is the energy of single photon at $\lambda=\SI{780}{\nm}$ and the detecting time length $\tau=\SI{109.08}{\ms}$ in our experiment. Thus, the effective measured photons number is fixed as $N=\num{4.04d7}$ in this experiment. 

Then exerting the driving signal on PZT chips and projecting the final pointer to $|\psi_{\hat{L}}\rangle$, the detected photons number is
\begin{align}
	N_{\alpha} &= \left|\langle\psi_{\hat{L}}|\hat{\Pi}_{\hat{L}}|\psi_{\hat{L}}\rangle\right|^{2}N = (2mn+m+n)(\cot\varepsilon)^{2}\alpha_{tot}^{2}N \nonumber\\
	&\approx (2mn+m+n)(\cot\varepsilon)^{2}\alpha_{0}^{2}N+2(2mn+m+n)(\cot\varepsilon)^{2}\alpha_{0}\alpha\cos(2\pi ft)N \label{eq:23}
\end{align}
Similarly, we have the detected power of APD is
\begin{align}
	I_{\alpha} &\approx (2mn+m+n)(\cot\varepsilon)^{2}\alpha_{0}^{2}I_{0} \nonumber \\
	&\quad+2(2mn+m+n)(\cot\varepsilon)^{2}\alpha_{0}\alpha\cos(2\pi ft)I_{0} \label{eq:24}
\end{align}
Inputting the detected power signal of APD into a spectrum analyzer, the tiny rotation signal at $f=\SI{1}{\kHz}$ is demodulated as:
\begin{equation}
	I_{\alpha}^{\SI{1}{\kHz}} = 2(2mn+m+n)(\cot\varepsilon)^{2}\alpha_{0}\alpha I_{0} \label{eq:25}
\end{equation}
From Eq. \ref{eq:23}, we can obtain the shot-noise of APD is $\delta N_{\alpha}=\sqrt{N_{\alpha}}\approx\sqrt{2mn+m+n}|\cot\varepsilon|\alpha_{0}\sqrt{N}$, and the corresponding shot-noise power is
\begin{equation}
	\delta I_{\alpha} = \gamma \delta N_{\alpha}/\tau \approx \gamma\sqrt{2mn+m+n}|\cot\varepsilon|\alpha_{0}\sqrt{N}/\tau \label{eq:26}
\end{equation}
Thus, the detected peak signal-to-noise ratio of spectrum analyzer is
\begin{equation}
	\mathrm{SNR} = \frac{I_{\alpha}^{\SI{1}{\kHz}}}{\delta I_{\alpha}} = 2\sqrt{2mn+m+n}|\cot\varepsilon|\sqrt{N}\alpha \label{eq:27}
\end{equation}
When $\mathrm{SNR}=1$, the minimum detectable rotation signal can be obtained as:
\begin{equation}
	\alpha_{\min}^{(m,n)} = \frac{1}{\sqrt{2mn+m+n}} \frac{1}{2|\cot\varepsilon|\sqrt{N}} \label{eq:28}
\end{equation}

\begin{figure}[h]
	\centering
	\includegraphics[width=0.8\linewidth]{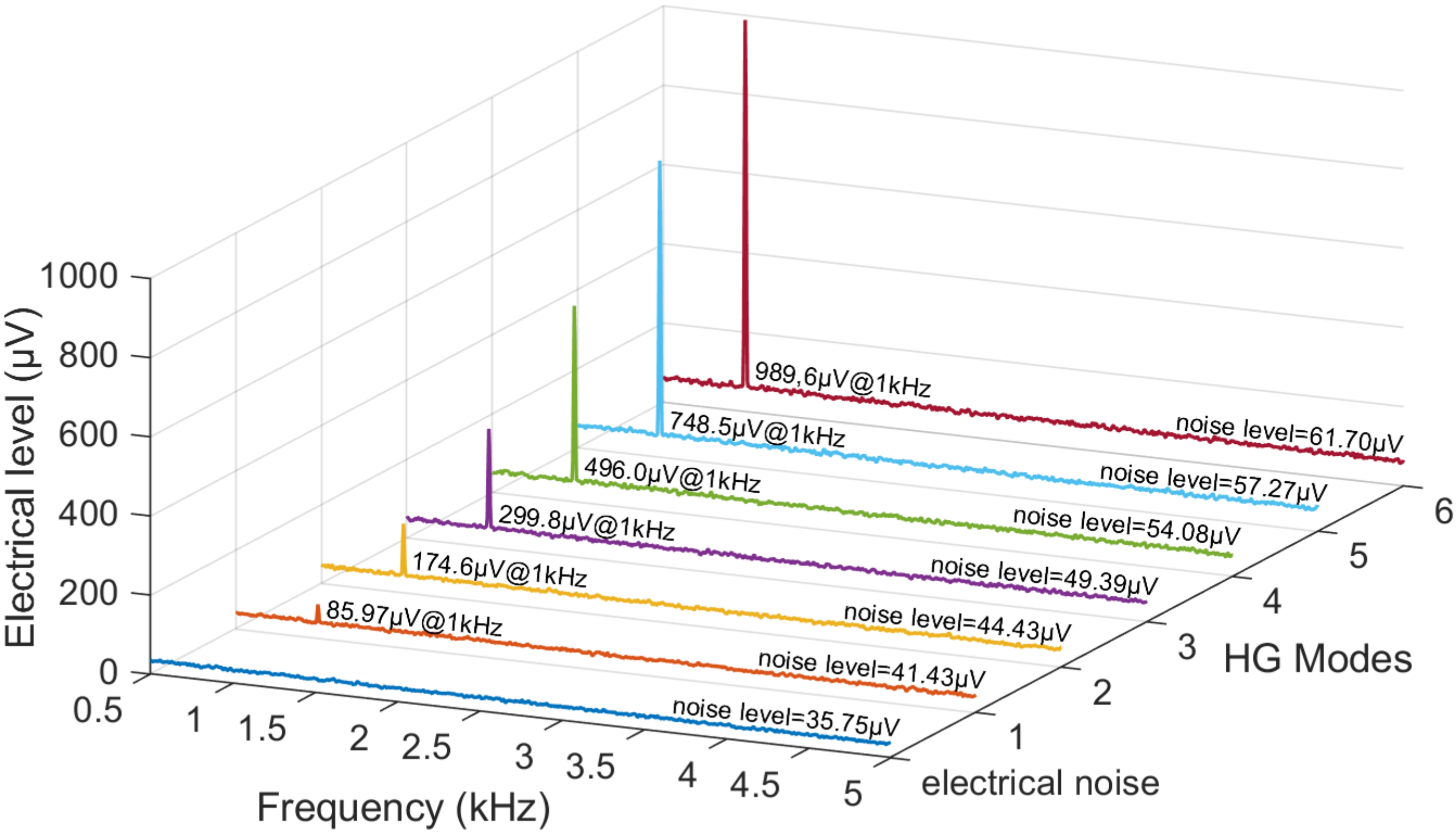}
	\caption{\label{fig:4} Detected electrical spectrum of HG11 to HG66 modes at $\SI{500}{\Hz}$ to $\SI{5}{\kHz}$. Driving voltage of PZT is $\SI{5}{\V}$, corresponds to $\SI{22}{\micro rad}$ rotation signal. The first line is the spectrum of electrical noise floor of APD detector, which is detected without input light on APD.}
\end{figure}

In practice, we detected the peak level from spectrum analyzer at $\SI{1}{\kHz}$ to demodulate the amplitude of rotation signal. Generally, the peak level consists of three parts: signal level, shot-noise floor and electrical noise floor, which is denoted as $V_{p}=V_{\alpha}+V_{sn}+V_{en}$. Here, $V_{\alpha}\propto I_{\alpha}^{(\SI{1}{\kHz})}$ is the signal level, $V_{sn}\propto\delta I_{\alpha}$ is the shot noise level, they vary with the different HG modes. In our experiment, the electrical noise level $V_{en}=\SI{35.75}{\uV}$ is a constant value in the experiment, which was detected without inputting light on the APD. The detected level of total noise floor with $mn$-order HG beam is $V_{noise}^{(m,n)}=V_{sn}^{(m,n)}+V_{en}$. Thus, the detected signal-to-noise ratio with $mn$-order HG beam in our scheme is obtained as:
\begin{equation}
	\mathrm{SNR}^{(m,n)} = \frac{V_{\alpha}^{(m,n)}}{V_{sn}^{(m,n)}} = \frac{V_{p}^{(m,n)}-V_{noise}^{(m,n)}}{V_{noise}^{(m,n)}-V_{en}} \label{eq:29}
\end{equation}
To determine the detected noise levels, we illustrate the electrical spectrums of HG11 to HG66 modes at $\SI{500}{\Hz}$ to $\SI{5}{\kHz}$ with Driving voltage $\SI{5}{\V}$ in Fig. \ref{fig:4}.

As is shown in Fig. \ref{fig:4}, the electrical noise is $V_{en}=\SI{35.75}{\uV}$, and the detected shot-noise level of $m,n$-order HG beam can be calculated by $V_{sn}^{(m,n)}=V_{noise}^{(m,n)}-V_{en}$. Here, we list the results in Tab. \ref{tab:1}.
\begin{table}[htb]
	\begin{minipage}{\linewidth}
		\centering
		\caption{Experimental results of detected noise levels.}
		\label{tab:1}
		\begin{threeparttable}
			\begin{tabular*}{\linewidth}{ccccccc}
				\toprule
				HG mode & HG11 & HG22 & HG33 & HG44 & HG55 & HG66 \\
				\midrule
				noise level\tnote{\textit{a}} & $\SI{41.43}{\uV}$ & $\SI{44.43}{\uV}$ & $\SI{49.39}{\uV}$ & $\SI{54.08}{\uV}$ & $\SI{57.27}{\uV}$ & $\SI{61.70}{\uV}$ \\
				shot-noise level & $\SI{5.68}{\uV}$ & $\SI{8.68}{\uV}$ & $\SI{13.64}{\uV}$ & $\SI{18.33}{\uV}$ & $\SI{21.52}{\uV}$ & $\SI{25.95}{\uV}$ \\
				\bottomrule
			\end{tabular*}
			\begin{tablenotes}
				\footnotesize
				\item[\textit{a}] Total detected noise levels in the APD.
			\end{tablenotes}
		\end{threeparttable}
	\end{minipage}
\end{table}

\begin{figure*}[htb]
	\centering
	\begin{minipage}{0.49\linewidth}
		\centering
		\begin{overpic}[width=\linewidth]{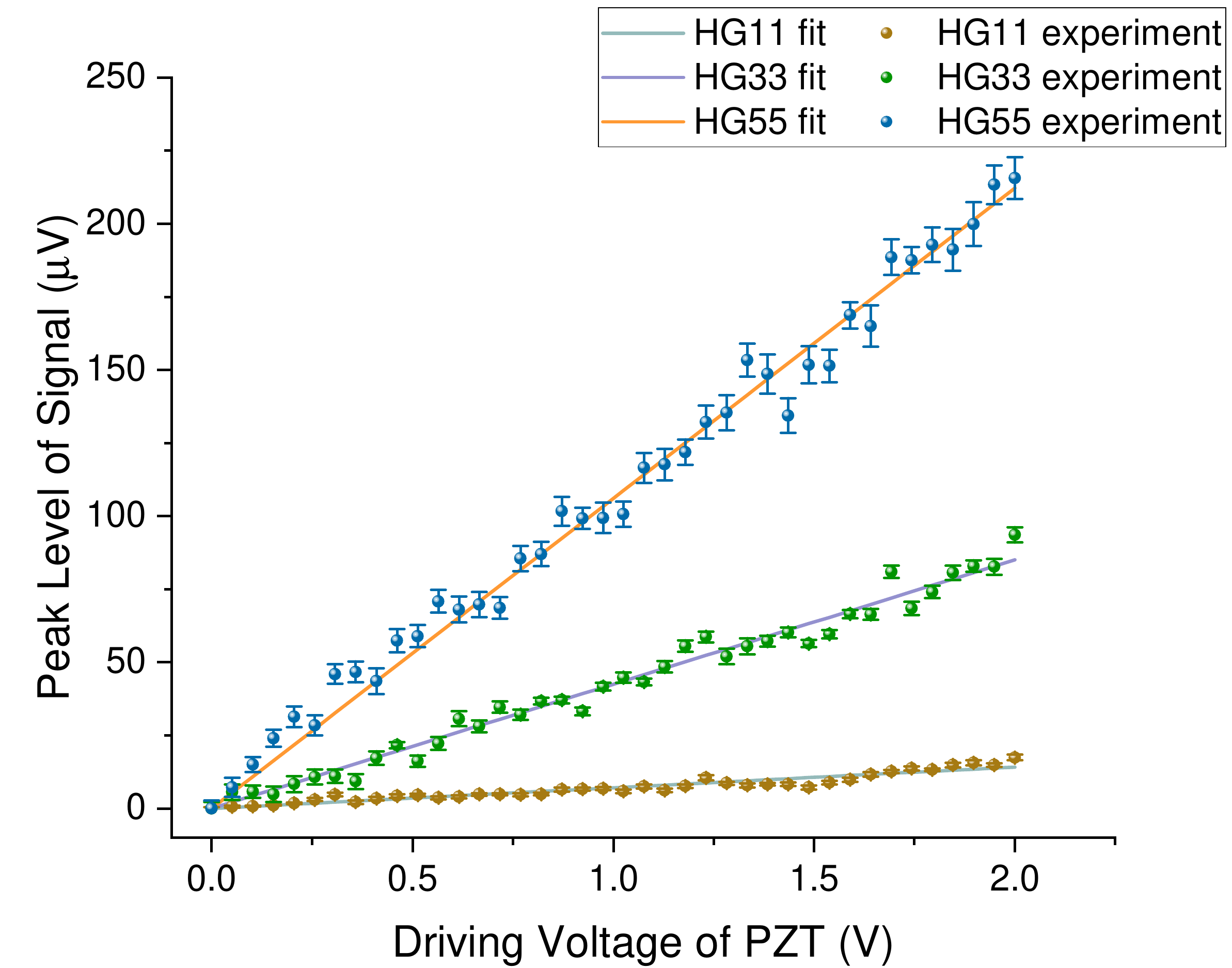}
			\put(1,75){(\textsf{a})}
		\end{overpic}
	\end{minipage}
	\begin{minipage}{0.49\linewidth}
		\centering
		\begin{overpic}[width=\linewidth]{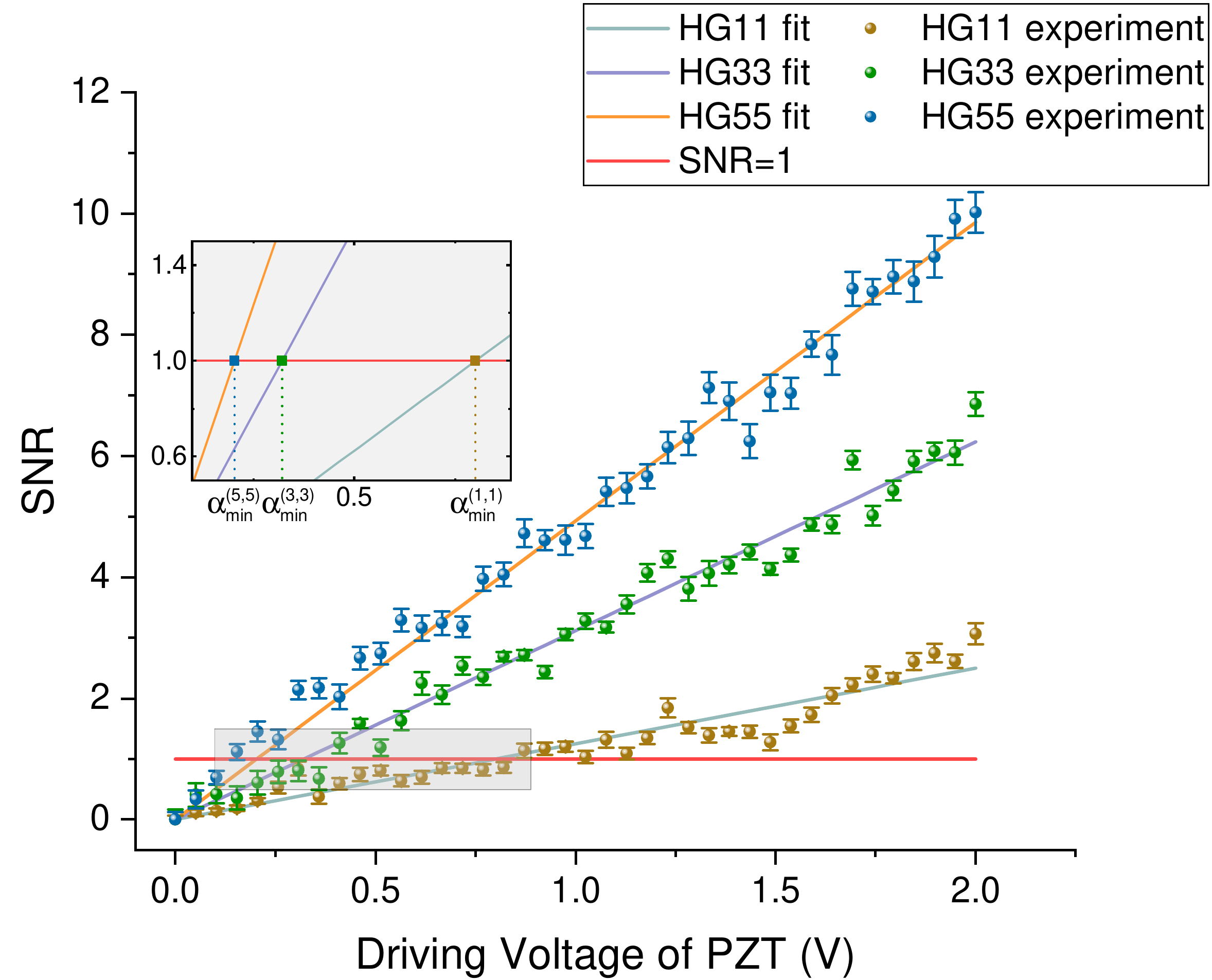}
			\put(1,75){(\textsf{b})}
		\end{overpic}
	\end{minipage}
	\caption{\label{fig:5} Experimental results. (a) Detected peak signal level of HG11, HG33 and HG55 modes at 1kHz. (b) Detected signal-to-noise ratio of HG11, HG33 and HG55 modes. Driving voltage of PZT increases from $\SI{0}{\V}$ to $\SI{2}{\V}$, corresponds to $\SI{0}{\micro rad}$ to $\SI{8.8}{\micro rad}$ rotation signal.}
\end{figure*}

In Fig. \ref{fig:5}, we illustrate the experimental results of detected peak signal level and signal-to-noise ratio at $1\mathrm{kHz}$ with 1$\times$1-order, 3$\times$3-order and 5$\times$5-order Hermite-Gaussian modes. Finally, a significant precision of $\SI{0.89}{\micro rad}$ is achieved with HG55 mode in this experiment. Here, we list the experimental results of the driving voltages of PZT at $\mathrm{SNR}=1$ and the corresponding minimal detected rotation angles in Tab. \ref{tab:2}. For comparison, we also calculate the theoretical predictions of minimal detectable rotations based on Eq. \ref{eq:28} with fixed detected photon number $N=\num{4.04d7}$ and post-selection angle $\varepsilon=\ang{5}$. Our experiment results is consistent well with the theoretical predictions.

\begin{table}[htb]
	\begin{minipage}{\linewidth}
		\centering
		\caption{Minimal detectable rotation angles with different HG modes.}
		\label{tab:2}
		\begin{threeparttable}
			\begin{tabular*}{\linewidth}{@{\extracolsep{\fill}}cccc@{\extracolsep{\fill}}}
				\toprule
				& Theory\tnote{\textit{a}} & \multicolumn{2}{@{}c@{}}{Experiment} \\ \cmidrule(lr){2-2} \cmidrule(lr){3-4}
				HG modes & $\alpha_{\min}^{\mathrm{th}}$ & $V_{\mathrm{PZT}}$\tnote{\textit{b}} & $\alpha_{\min}^{\mathrm{exp}}$ \\
				\midrule
				HG11 & $\SI{3.44}{\micro rad}$ & $\SI{801}{\mV}$ & $\SI{3.52}{\micro rad}$ \\
				HG33 & $\SI{1.40}{\micro rad}$ & $\SI{321}{\mV}$ & $\SI{1.41}{\micro rad}$ \\
				HG55 & $\SI{0.89}{\micro rad}$ & $\SI{203}{\mV}$ & $\SI{0.89}{\micro rad}$ \\
				\bottomrule
			\end{tabular*}
			\begin{tablenotes}
				\footnotesize
				\item[\textit{a}] These theoretical predictions are derived from Eq. \ref{eq:28} with fixed detected photon number $N=\num{4.04d7}$ and post-selection angle $\varepsilon=\ang{5}$.
				\item[\textit{b}] The driving voltages of PZT chips at $\mathrm{SNR}=1$.
			\end{tablenotes}
		\end{threeparttable}
	\end{minipage}
\end{table}

\section{Discussions}
\subsection{Technical advantages of weak value}
In the theoretical frame, the post-selection is employed for individually reading out the measurement parameters from pointer, and the precision enhancement comes from the mode entanglement of HG pointer, but not the weak values. Thus, our main conclusion still holds in the post-selection-free scheme. In the experimental scheme, we still employed the weak value amplification technology. Though the weak value $A_{w}=\cot\varepsilon$ takes no enhancement for the theoretical minimum detectable rotation in Eq. \ref{eq:28} because the detected photons' number $N=\left|\langle f|i\rangle\right|^{2}N_{0}=\sin^{2}\varepsilon N_{0}$ is attenuated by the post-selection, where $N_{0}$ is the photons number before post-selection. However, the weak value amplification technology has been proved efficient for suppressing technical noises, such as reflection of optical elements\cite{PhysRevA.97.063818} and detector saturation\cite{Xu:18,PhysRevLett.125.080501}. Especially, the detector saturation is non-negligible in our experiment for the saturation power of our APD detector is only $\SI{1.54}{\nano\W}$. Considering the projection demodulation of SLM, only $~10\%$ photons can be modulated on the 1st-order diffraction, so the maximum efficient received power of our detector is about $\SI{154}{\pico\W}$, which is easily saturated without post-selection. For example, the efficient detected light power in our experiment is $I_{0}=\SI{94.34}{\pico\W}$, and the post-selected angle $\varepsilon=\ang{5}$. Therefore, for the post-selection-free scheme, a $I_{0}/\sin^{2}\varepsilon=\SI{12.42}{\nano\W}$ detected light power is needed to achieve the same precision of the post-selected scheme, which is far lager than the saturation power of the APD detector.

\subsection{Rotation-coupling weak measurement for Hamiltonian estimation}
Though we only investigate the enhanced measurement on angular rotation, the precision enhancement of employing HG pointers can be applied in various missions in quantum physics. The most obvious application of our scheme in quantum physics is the estimation of Hamiltonian\cite{PhysRevA.95.022335,10.1088/1361-6455/abe5c7}. In this case, we do not only concentrate on the interaction strength parameter $\alpha$, but also interest in the information of operator $\hat{A}$. For two-level system, the unknown operator can be represented as $\hat{A}=\vec{n}\cdot\vec{\sigma}$, where $\vec{n}=\left(\sin\theta\cos\phi,\sin\theta\sin\phi,\cos\theta\right)$ is the direction vector of measurement operator, and $\vec{\sigma}=\left(\hat{\sigma}_{x},\hat{\sigma}_{y},\hat{\sigma}_{z}\right)$ where $\hat{\sigma}_{x}$, $\hat{\sigma}_{y}$, and $\hat{\sigma}_{z}$ are Pauli matrices. Thus, there are two unknown parameters $\theta$ and $\phi$ to be estimated for identifying the operator $\hat{A}$. As we calculated in the theoretical model, the final pointer's state in our post-selected scheme is $|\psi_{f}\rangle\approx\mathcal{N}\left(1-\mathrm{i}\alpha A_{w}\hat{\Omega}\right)|\psi_{i}\rangle$. Then the QFI of parameter $g\in (\theta,\phi)$ can be calculated as $\mathcal{Q}(g)\approx 4\alpha^{2}\left|\partial_{g}A_{w}\right|^{2}\langle\Delta\hat{\Omega}^{2}\rangle_{i}$. Combining with the QCR inequality $\delta\hat{g}^{2}\ge 1/N\mathcal{Q}(g)$, the estimation precision of parameter $g\in (\theta,\phi)$ satisfies the uncertainty relation:
\begin{equation}
	\delta\hat{g}^{2}\langle\Delta\hat{\Omega}^{2}\rangle_{i} \ge \frac{1}{4N\alpha^{2}\left|\partial_{g}A_{w}\right|^{2}} \label{eq:30}
\end{equation}
where the quantum limits of Hamiltonian parameters are still governed by the variance of translation operator on the initial pointer. It means that our scheme has potential to be applied in this scenario for improving the performance of Hamiltonian estimation. For example, the initial pointer with Gaussian profile are employed traditionally, and the two-level system couples with the pointer via a displacement interaction, $\hat{\Omega}=\hat{p}_{x}$. Then the variances $\langle\Delta\hat{\Omega}^{2}\rangle_{i}=\langle\Delta\hat{p}_{x}^{2}\rangle_{\mathrm{G}}=1/4\sigma_{0}$, and the corresponding quantum limit on estimating Hamiltonian parameters $g\in (\theta,\phi)$ is given by:
\begin{equation}
	\delta\hat{g}^{2} \ge \frac{\sigma_{0}^{2}}{N\alpha^{2}\left|\partial_{g}A_{w}\right|^{2}} \label{eq:31}
\end{equation}
If we replace the Gaussian pointer by $mn$-order HG pointer, the quantum limit will be improved as:
\begin{equation}
	\delta\hat{g}^{2} \ge \frac{\sigma_{0}^{2}}{(2m+1)N\alpha^{2}\left|\partial_{g}A_{w}\right|^{2}} \label{eq:32}
\end{equation}
because the variance $\langle\Delta\hat{p}_{x}^{2}\rangle_{mn}=(2m+1)/4\sigma_{0}$ increases linearly with the HG mode $m$ in the corresponding displacement direction. Further, replacing the displacement interaction by rotational interaction, that is $\hat{\Omega}=\hat{L}_{z}$, there will be a significant improvement on the precision limit on estimating Hamiltonian parameters:
\begin{equation}
	\delta\hat{g}^{2} \ge \frac{1}{4(2mn+m+n)N\alpha^{2}\left|\partial_{g}A_{w}\right|^{2}} \label{eq:33}
\end{equation}
which is quadratically improved by the HG mode numbers $m$ and $n$. Moreover, the enhancement factor $2mn+m+n$ is analog to the Heisenberg scaling limit in quantum interference\cite{doi:10.1116/5.0062114}. Because the mode number $m$ in $x$-direction and mode number $n$ in $y$-direction are independent for HG beam state or 2-D HO state. Thus, mode state $|m,n\rangle$ can be regarded as an eigenket in the product Hilbert space $\mathcal{H}_{x}\otimes\mathcal{H}_{y}$, where $\mathcal{H}_{x}$ and $\mathcal{H}_{y}$ are the Hilbert spaces for mode state in $x$-direction and mode state in $y$-direction. Therefore, the OAM operator $\hat{L}_{z}=\mathrm{i}\left(\hat{a}_{x}\hat{a}_{y}^{\dagger}-\hat{a}_{x}^{\dagger}\hat{a}_{y}\right)$ in this product Hilbert space $\mathcal{H}_{x}\otimes\mathcal{H}_{y}$, which leads to the unknown parameters $(\alpha,\theta,\phi)$ taken by the state $|\psi_{\hat{L}}\rangle$. Moreover, it is obvious to note that the state $|\psi_{\hat{L}}\rangle$ in Eq. \ref{eq:20} is a mode-entangled state in the product Hilbert space $\mathcal{H}_{x}\otimes\mathcal{H}_{y}$.

\subsection{Rotation-coupling weak measurement for monitoring quantum bit}
Besides Hamiltonian estimation, our precision-enhanced method also has potential for monitoring the quantum bit, which is a vital mission in quantum metrology\cite{Minev_2019}. Generally, the state of an arbitrary quantum bit (qubit) can be represented as:
\begin{equation}
	|qubit\rangle = \cos\frac{\theta}{2}|0\rangle+\mathrm{e}^{\mathrm{i}\phi}\sin\frac{\theta}{2}|1\rangle \label{eq:34}
\end{equation}
where $\theta$ and $\phi$ are the azimuthal angles on the Bloch sphere, $|0\rangle$ and $|1\rangle$ are the eigenkets of Pauli operator $\hat{\sigma}_{z}$. To avoid apparent disturbance on the qubit, a series of continuous weak measurements is adopted to monitor it\cite{Hays_2020}.
\begin{figure}[h]
	\centering
	\includegraphics[width=0.5\linewidth]{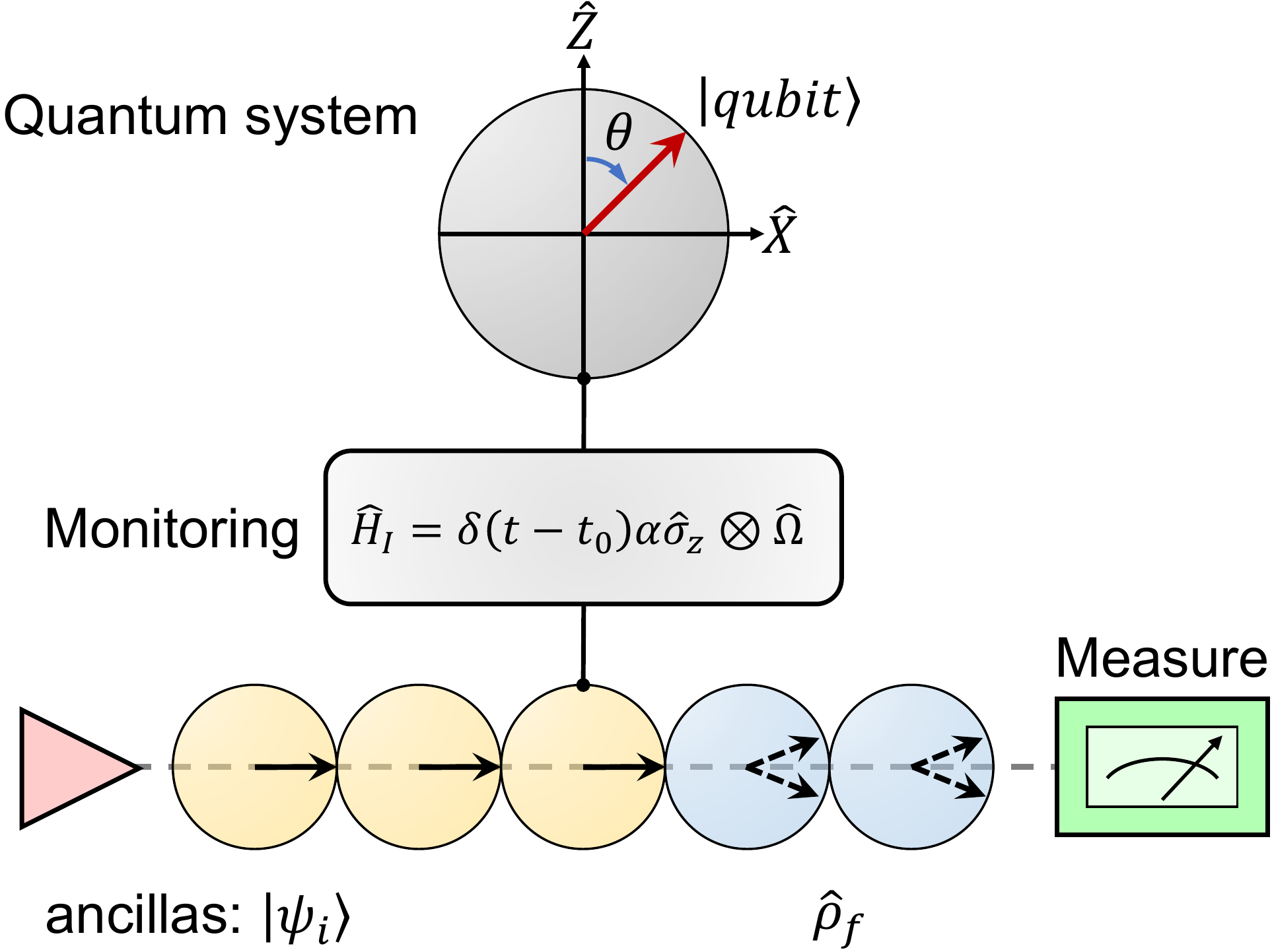}
	\caption{\label{fig:6} Schematic of monitoring quantum bit.}
\end{figure}

As is illustrated in Fig. \ref{fig:6}, an ancillary device (pointer) is employed to monitor the quantum system via an weak interaction procedure, which is described by the von Neumann measurement theory\cite{PhysRevLett.60.1351} via an impulse Hamiltonian (here we take the monitoring of qubit on the bases of $\hat{\sigma}_{z}$ for example):
\begin{equation}
	\hat{H}_{I} = \delta(t-t_0)\alpha\hat{\sigma}_{z}\otimes\hat{\Omega} \label{eq:35}
\end{equation}
which leads to a composite unitary evolution $\hat{U}=\exp\left(-\mathrm{i}\int\hat{H}_{I}\mathrm{d}t\right)\approx 1-\mathrm{i}\alpha\hat{\sigma}_{z}\otimes\hat{\Omega}$ of quantum system and pointer. After the weak interaction, the measurement information of Pauli operator $\hat{\sigma}_{z}$ on the qubit is transferred to the pointer shift of ancillas via a translation operator $\hat{\Omega}$, and the final state of whole system is
\begin{equation}
	|\Psi_{f}\rangle = \cos\frac{\theta}{2}|0\rangle|\psi_{+}\rangle+\mathrm{e}^{\mathrm{i}\phi}\sin\frac{\theta}{2}|1\rangle|\psi_{-}\rangle \label{eq:36}
\end{equation}
where $|\psi_{+}\rangle=\exp(-\mathrm{i}\alpha\hat{\Omega})|\psi_{i}\rangle$ and $|\psi_{-}\rangle=\exp(\mathrm{i}\alpha\hat{\Omega})|\psi_{i}\rangle$. Unlike our post-selected weak measurement scheme, post-selection of qubit is forbidden. Thus, we should measure the information about parameter $\theta$ from the final pointer state, which is a mixed state:
\begin{equation}
	\hat{\rho}_{f} = \cos^{2}\frac{\theta}{2}|\psi_{+}\rangle\langle\psi_{+}|+\sin^{2}\frac{\theta}{2}|\psi_{-}\rangle\langle\psi_{-}| \label{eq:37}
\end{equation}
then the corresponding QFI of parameter $\theta$ can be calculated as $\mathcal{Q}(\theta)=1-\left(\mathrm{Re}\langle\psi_{+}|\psi_{-}\rangle\right)^{2}\approx 4\alpha^{2}\langle\hat{\Omega}^{2}\rangle_{i}$, which leads to the quantum limit on measuring parameter $\theta$ be governed by the uncertainty relation (See the Appendix \ref{sec:A2} for derivation)
\begin{equation}
	\delta\hat{\theta}^{2}\langle\hat{\Omega}^{2}\rangle_{i} \ge 
	\frac{1}{4\alpha^{2}N} \label{eq:38}
\end{equation}
which means that the monitoring sensitivity is still dependent on the devising of pointer and coupling method. Applying our $mn$-order HG pointer and rotation-coupling method to this scheme, the quantum limit on monitoring the azimuthal angle $\theta$ of qubit is then derived as:
\begin{equation}
	\delta\hat{\theta}^{2} \ge 
	\frac{1}{4\alpha^{2}(2mn+m+n)N} \label{eq:39}
\end{equation}
where the precision enhancement still holds.

Moreover, we express the HG beams via harmonic oscillator model, which has been widely used in quantum computation and metrology, such as superconducting qubits\cite{LaHaye_2009,doi:10.1146/annurev-conmatphys-031119-050605} and optomechanics\cite{Connell_2010,PhysRevLett.109.023601,Balram_2016}. Thus, our theoretical model can be applied in such scenarios naturally, and provide a significant method in quantum metrology.

\section{Conclusions}
In summary, we have implemented a practical scheme for measuring the tiny rotation by employing $mn$-order HG pointer in a post-selected weak measurement scheme, where the precision limit is improved by a factor of $2mn+m+n$ theoretically. Experimentally, we demodulate the angular rotation parameter via a single projective projective measurement, and precision up to $\SI{0.89}{\micro rad}$ is achieved with $5\times 5$-order HG beams. Moreover, we have found that the precision enhancement of rotation-coupling method with HG pointer still holds in a wide range of applications in quantum physics, such as Hamiltonian estimation and monitoring qubit. Thus, our results constitute valuable resources not for measurement and controlling of light’s angular rotation in optical metrology, but also for sensitive estimating and controlling of evolution procedure in quantum physics.

\appendix
\section{Derivation of quantum limits in post-selected weak measurement scheme} \label{sec:A1}
In this section, we give the calculation details about the quantum limits of weak interaction parameters $\alpha$, $\theta$ and $\phi$ in post-selection weak measurement, where $\alpha$ is the weak interaction strength and $\alpha\ll 1$, $\theta$ and $\phi$ are the Hamiltonian parameters of the two-level system. In this case, the operator $\hat{A}$ is given as the generalized formalism $\hat{A}=\vec{n}\cdot\vec{\sigma}$, $\vec{\sigma}=\left(\hat{\sigma}_{x},\hat{\sigma}_{y},\hat{\sigma}_{z}\right)$, where $\hat{\sigma}_{x}$, $\hat{\sigma}_{y}$ and $\hat{\sigma}_{z}$ are the Pauli matrices. The weak interaction procedure is described by an impulse Hamiltonian is $\hat{H}_{I}=\delta\left(t-t_{0}\right)\alpha\hat{A}\otimes\hat{\Omega}$, then the evolution operator of weak interaction procedure can be calculated as:
\begin{align}
	\hat{U} &= \exp\left(-\mathrm{i}\hat{H}_{I}\mathrm{d}t\right) = \exp\left(-\mathrm{i}\alpha\hat{A}\otimes\hat{\Omega}\right) \nonumber\\
	&= \frac{1}{2}\left(\hat{I}+\vec{n}\cdot\vec{\sigma}\right)\exp\left(-\mathrm{i}\alpha\hat{\Omega}\right)+\frac{1}{2}\left(\hat{I}-\vec{n}\cdot\vec{\sigma}\right)\exp\left(\mathrm{i}\alpha\hat{\Omega}\right) \nonumber\\
	&\approx 1-\mathrm{i}\alpha\vec{n}\cdot\vec{\sigma}\otimes\hat{\Omega} \label{eq:A1}
\end{align}

The initial state of whole system before weak interaction can be denoted as $|\Psi_{i}\rangle=|\psi_{i}\rangle|i\rangle$. Then the final state of whole system after weak interaction and post-selection can be calculated as:
\begin{align}
	|\tilde{\Psi}_{f}\rangle &= |f\rangle\langle f|\hat{U}|\Psi_{i}\rangle \approx |f\rangle\langle f|\left(1-\mathrm{i}\alpha\vec{n}\cdot\vec{\sigma}\otimes\hat{\Omega}\right)|\Psi_{i}\rangle \nonumber\\
	&= \left[\langle f|i\rangle\left(1-\mathrm{i}M_{w}\hat{\Omega}\right)|\psi_{i}\rangle\right]\otimes|f\rangle \label{eq:A2}
\end{align}
Here we denote $\alpha A_{w}=M_{w}$ for simplicity. Then the pointer's final state is expressed as $|\tilde{\psi}_{f}\rangle=\langle f|i\rangle\left(1-\frac{i}{\hbar}M_{w}\hat{\Omega}\right)|\psi_{i}\rangle$, which can be normalized as:
\begin{equation}
	|\psi_{f}\rangle = \mathcal{N}\left(1-\mathrm{i}M_{w}\hat{\Omega}\right)|\psi_{i}\rangle \label{eq:A3}
\end{equation}
where
\begin{equation}
	\mathcal{N} = \frac{1}{\sqrt{1+2\mathrm{Im}\left(M_{w}\right)\langle\hat{\Omega}\rangle_{i}+\left|M_{w}\right|^{2}\langle\hat{\Omega}^{2}\rangle_{i}}} \label{eq:A4}
\end{equation}
is the normalized factor.

For a parameterized state $|\psi(g)\rangle$, its corresponding QFI for a single parameter $g$ can be given by\cite{DEMKOWICZDOBRZANSKI2015345,Liu_2019}:
\begin{equation}
	\mathcal{Q}\left(g\right) = 4\left(\frac{\partial\langle\psi(g)|}{\partial g}\frac{\partial|\psi(g)\rangle}{\partial g}-\frac{\partial\langle\psi(g)|}{\partial g}|\psi(g)\rangle\langle\psi(g)|\frac{\partial|\psi(g)\rangle}{\partial g}\right) \label{eq:A5}
\end{equation}
Substituting $|\psi_{f}\rangle$ into Eq. \ref{eq:A5}, the QFI of each measurement parameter can be calculated as:
\begin{align}
	\mathcal{Q}\left(g\right) &= 4\mathcal{N}^{2}\left[\left|\frac{\partial M_{w}}{\partial g}\right|^{2}\langle\hat{\Omega}^{2}\rangle_{i}-\mathcal{N}^{2}\left(\left|\frac{\partial M_{w}}{\partial g}\right|^{2}\langle\hat{\Omega}\rangle_{i}^{2}\right.\right. \nonumber\\
	&\left.\left.\qquad\qquad+\mathrm{Im}M_{w}\left|\frac{\partial M_{w}}{\partial g}\right|^{2}\langle\hat{\Omega}\rangle_{i}\langle\hat{\Omega}^{2}\rangle_{i}+\left|M_{w}\right|^{2}\left|\frac{\partial M_{w}}{\partial g}\right|^{2}\langle\hat{\Omega}^{2}\rangle_{i}^{2}\right)\right] \nonumber\\
	&\approx 4\left|\frac{\partial M_{w}}{\partial g}\right|^{2}\langle\Delta\hat{\Omega}^{2}\rangle_{i} \label{eq:A6}
\end{align}
where $g\in\left(\alpha,\theta,\phi\right)$. Utilizing the quantum Cram$\acute{\mathrm{e}}$r-Rao (QCR) inequality $\delta\hat{g}^{2}\ge 1/N\mathcal{Q}\left(g\right)$, a coupling-parameter uncertainty relation can be obtained as:
\begin{equation}
	\delta\hat{g}^{2}\langle\Delta\hat{\Omega}^{2}\rangle_{i} \ge \frac{1}{4N}\cdot\frac{1}{\left|\partial_{g}M_{w}\right|^{2}} \label{eq:A7}
\end{equation}
Note that this result is derived under the approximate condition $M_{w}\ll 1$.

Taking $\vec{n}=\left(\sin\theta\cos\phi,\sin\theta\sin\phi,\cos\theta\right)$, we can calculate:
\begin{equation}
	\label{eq:A8}\left\{
	\begin{split}
		&\frac{\partial M_{w}}{\partial\alpha} = \sigma_{xw}\sin\theta\cos\phi+\sigma_{yw}\sin\theta\sin\phi+\sigma_{zw}\cos\theta \\
		&\frac{\partial M_{w}}{\partial\theta} = \alpha\left(\sigma_{xw}\cos\theta\cos\phi+\sigma_{yw}\cos\theta\sin\phi-\sigma_{zw}\sin\theta\right) \\
		&\frac{\partial M_{w}}{\partial\phi} = \alpha\left(\sigma_{yw}\sin\theta\cos\phi-\sigma_{xw}\sin\theta\sin\phi\right)
	\end{split}
	\right.
\end{equation}
where $\sigma_{xw}$, $\sigma_{yw}$ and $\sigma_{zw}$ are the corresponding weak values of Pauli operators $\hat{\sigma}_{x}$, $\hat{\sigma}_{y}$ and $\hat{\sigma}_{z}$ respectively. Then substituting Eq. \ref{eq:A6} and Eq. \ref{eq:A8} into uncertainty relation in Eq. \ref{eq:A7}, the lower bounds for $\delta\hat{\alpha}^{2}$, $\delta\hat{\theta}^{2}$ and $\delta\hat{\phi}^{2}$ can be calculated separately.
\begin{figure*}[htb]
	\centering
	\begin{minipage}{0.32\linewidth}
		\centering
		\begin{overpic}[width=\linewidth]{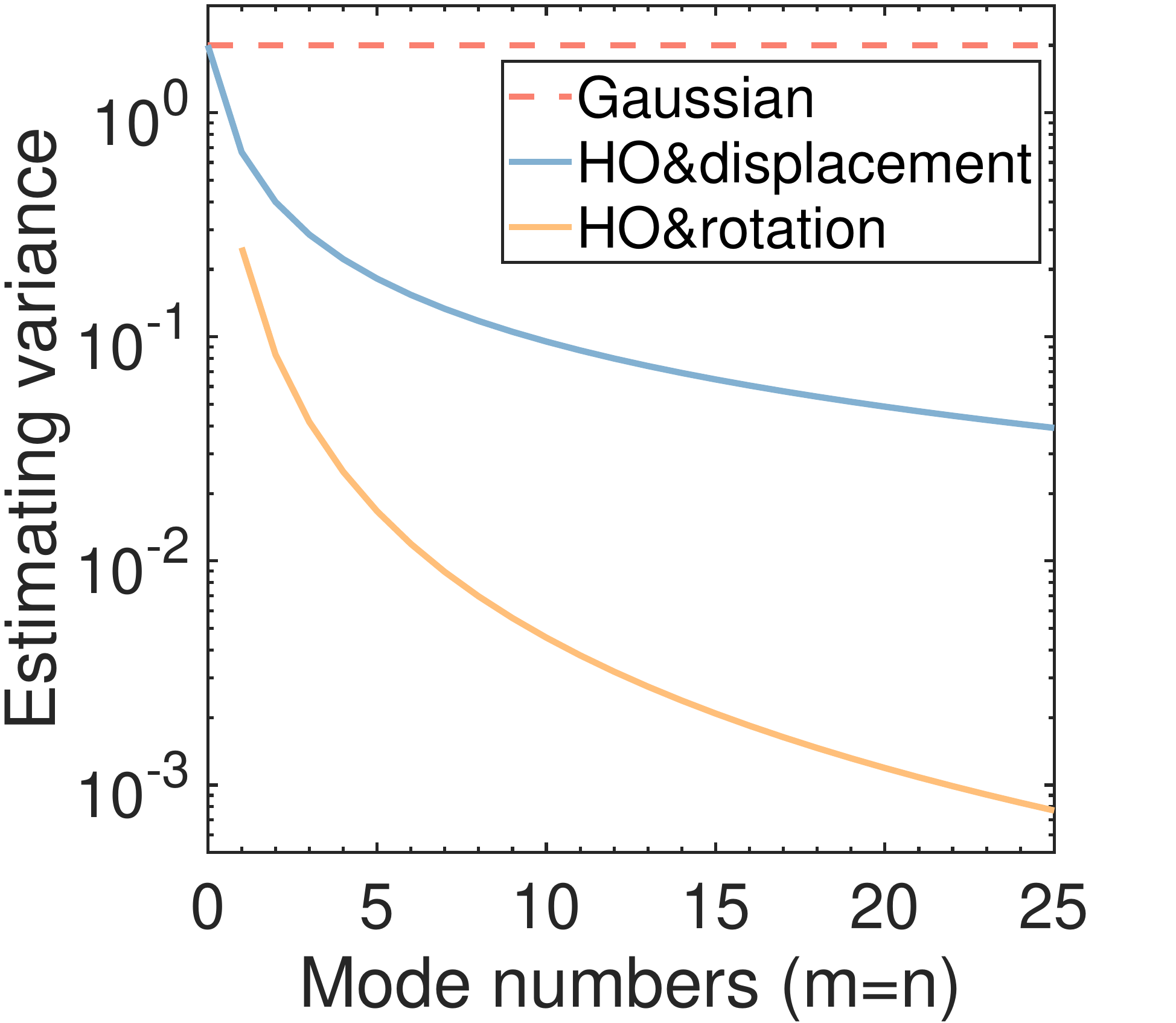}
			\put(3,85){(\textsf{a})}
		\end{overpic}
	\end{minipage}
	\begin{minipage}{0.32\linewidth}
		\centering
		\begin{overpic}[width=\linewidth]{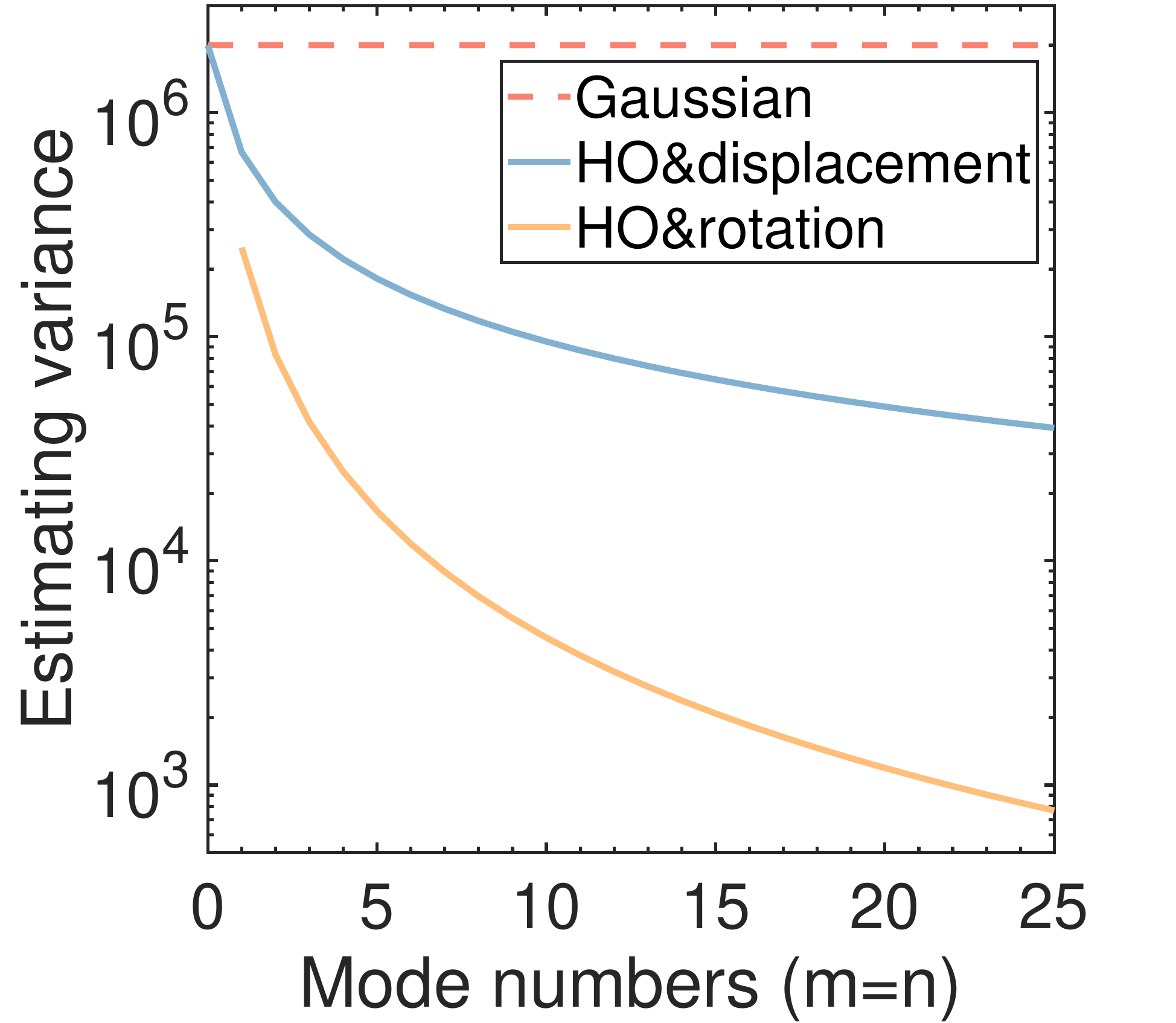}
			\put(3,85){(\textsf{b})}
		\end{overpic}
	\end{minipage}
	\begin{minipage}{0.32\linewidth}
		\centering
		\begin{overpic}[width=\linewidth]{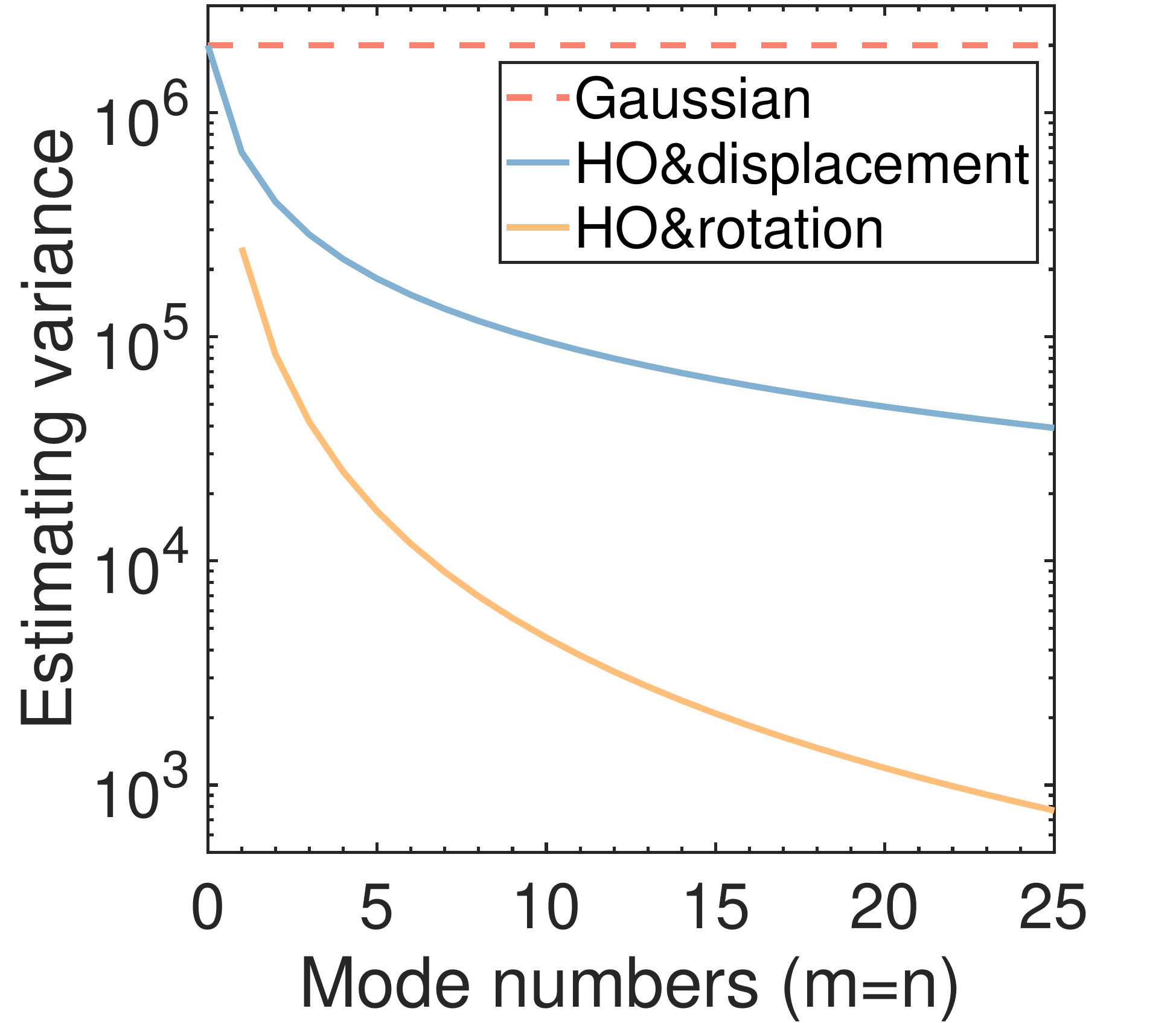}
			\put(3,85){(\textsf{c})}
		\end{overpic}
	\end{minipage}
	\caption{QCR bounds of the measurement parameters. (a) QCR bounds of parameter $\alpha$. (b) QCR bounds of parameter $\theta$. (c) QCR bounds of parameter $\phi$. The $y$-axis is the variance of estimator $\hat{g}$, and $x$-axis is the mode numbers. Mode numbers $m$ and $n$ simultaneously increase from 0 to 25. The red dotted line is the QCR bound of Gaussian pointer with displacement coupling. The blue line is the QCR bound of HO pointer with displacement coupling. The orange line is the QCR bound of HO pointer with rotation coupling. Here the pre-selection state and post-selection state are chosen as $|i\rangle=|f\rangle=\frac{1}{\sqrt{2}}\left(|0\rangle+\mathrm{e}^{\mathrm{i}\frac{\pi}{4}}|1\rangle\right)$, and values of every parameters are $\alpha=0.001$, $\theta=\pi/4$, $\phi=0$. Besides, the value of $\sigma_{0}$ is normalized to $1/\sqrt{2}$.}
	\label{fig:A1}
\end{figure*}

In this work, we investigate two types of pointer in the weak measurement scheme, Gaussian pointer and HG pointer. For Gaussian pointer, the measurement parameters are coupled to pointer's spatial displacement, which leads to a constant QCR bound. For HG pointer, its mode numbers in $x$-direction and $y$-direction are separately $m$ and $n$, the QCR bound is improved with pointer's mode numbers $m$ and $n$. Moreover, the 2-D HG pointer equals to a 2-D harmonic oscillator (HO) theoretically, which has been proved in the main text. Therefore, our results can be extended to any HO-formalism pointer here. Coupling the measurement parameters to the pointer's displacement ($x$-direction), the QCR bound is improved by factor $1/(2m+1)$, which is enhanced linearly. Coupling the measurement parameters to the pointer's rotation, the QCR bound is improved by factor $1/(2mn+m+n)$, which is enhanced quadratically. Here we calculate these three QCR bounds for parameters $\alpha$, $\theta$ and $\phi$ at $\alpha=0.001$, $\theta=\pi/4$ and $\phi=0$ separately. Without loss of generality, we normalize the value of pointer's spatial uncertainty $\sigma$ to $1/\sqrt{2}$, and set the measured samples number $N=1$. In Fig. \ref{fig:A1}, we illustrate the results for $\alpha$, $\theta$ and $\phi$ separately by choosing $|i\rangle=|f\rangle=\frac{1}{\sqrt{2}}\left(|0\rangle+e^{i\frac{\pi}{4}}|1\rangle\right)$, which equals to a part measurement of pointer without post-selection. Here $|0\rangle$ and $|1\rangle$ are the eigenkets of the Pauli operator $\hat{\sigma}_{z}$ on the two-level system.
\begin{figure}[h]
	\centering
	\includegraphics[width=0.8\linewidth]{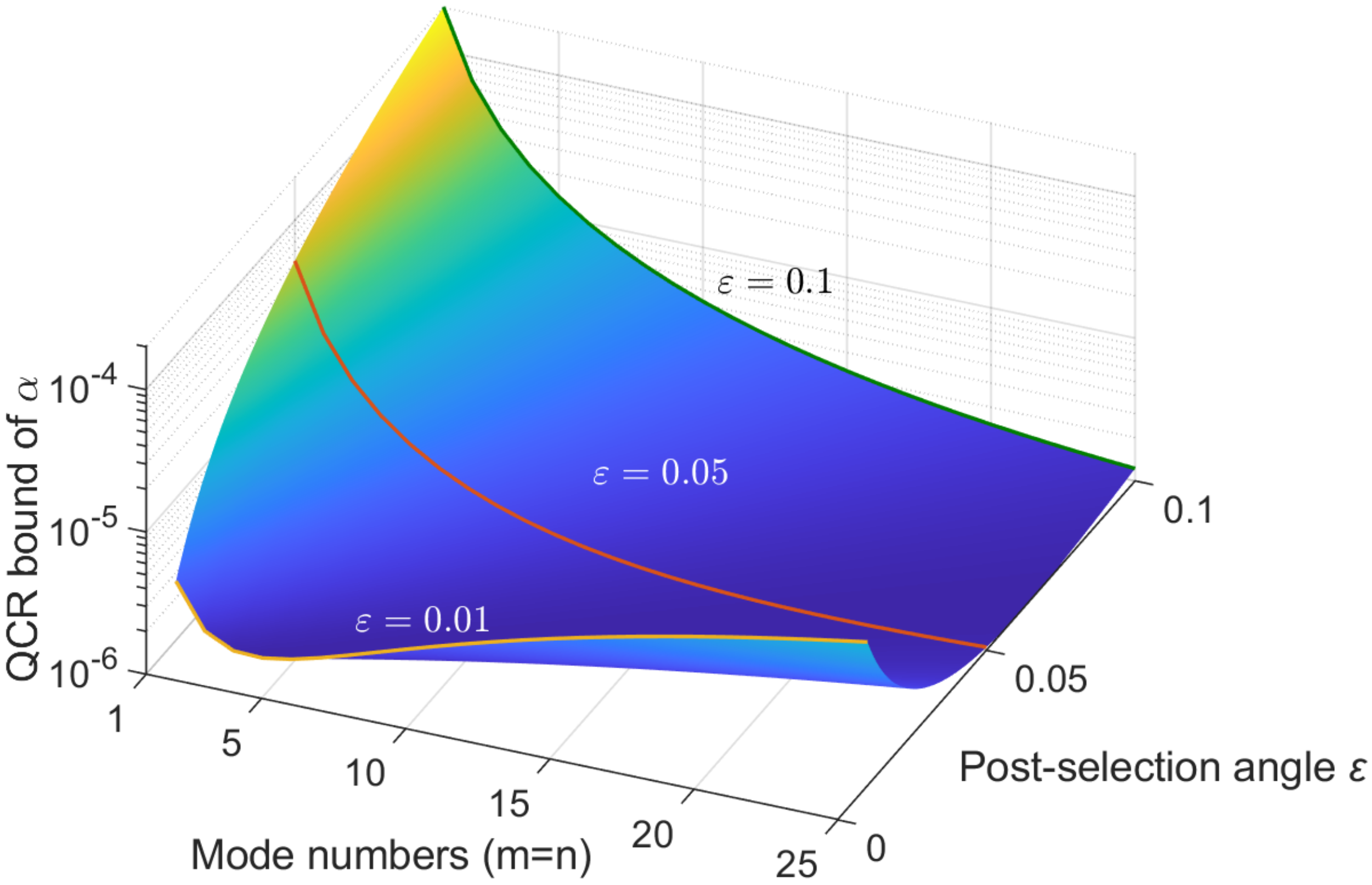}
	\caption{\label{fig:A2} QCR bound of $\alpha$ about post-selection angle $\varepsilon$. This is the QCR bound of $\alpha$ with 2-D HO pointer and rotation coupling. The $x$-axis is the mode numbers $m$ and $n$, which simultaneously increase from 1 to 25. Pre-selection state is $|i\rangle=\frac{1}{\sqrt{2}}\left(|0\rangle+e^{i\frac{\pi}{4}}|1\rangle\right)$ and post-selection state is $|f\rangle=\frac{1}{\sqrt{2}}\left(|0\rangle-e^{i\left(\frac{\pi}{4}+\varepsilon\right)}|1\rangle\right)$, where the angle $\varepsilon$ varies from 0.1 to 0.01. Besides, we plot three QCR bounds at different values of $\varepsilon$. The green line is $\varepsilon=0.1$, the red line is $\varepsilon=0.05$ and the yellow line is $\varepsilon=0.01$.}
\end{figure}

In practice, the post-selection state is usually nearly orthogonal to the pre-selection state in weak measurement scheme. Here we also analyze the impact of post-selection state's angle. Pre-selection state is still $|i\rangle=\frac{1}{\sqrt{2}}\left(|0\rangle+e^{i\frac{\pi}{4}}|1\rangle\right)$, and post-selection state is chosen as $|f\rangle=\frac{1}{\sqrt{2}}\left(|0\rangle-e^{i\left(\frac{\pi}{4}+\varepsilon\right)}|1\rangle\right)$, where $\varepsilon\ll 1$. To hold the approximate condition $M_{w}\ll 1$, we also require $\alpha/\varepsilon\ll 1$, otherwise the quadratic enhancement may vanishes. In Fig .\ref{fig:A2}, we illustrated the QCR bound of interaction strength $\alpha$ with HO pointer and rotation coupling, and $\varepsilon$ varies from 0.1 to 0.01. As is shown in Fig. \ref{fig:A2}, when the post-selection angle $\varepsilon$ is small enough to violate the approximate condition $M_{w}\ll 1$, the Heisenberg-like limit would vanish. Thus, the condition $M_{w}\ll 1$ should be strictly fulfilled in our scheme.

\section{Derivation of QFI for monitoring qubit with weak measurement scheme} \label{sec:A2}
As is elucidated in the main text, the ancillary pointer is adopted to monitor the qubit via a weak interaction procedure $\hat{U}\approx 1-\mathrm{i}\alpha\hat{\sigma}_{z}\otimes\hat{\Omega}$. Then the measurement information of Pauli operator $\hat{\sigma}_{z}$ on the qubit is transferred to the pointer shift of ancillas via a translation operator $\hat{\Omega}$, and the final state of whole system is
\begin{equation}
	|\Psi_{f}\rangle = \cos\frac{\theta}{2}|0\rangle|\psi_{+}\rangle+\mathrm{e}^{\mathrm{i}\phi}\sin\frac{\theta}{2}|1\rangle|\psi_{-}\rangle \label{eq:B1}
\end{equation}
where $|\psi_{+}\rangle=\exp(-\mathrm{i}\alpha\hat{\Omega})|\psi_{i}\rangle$ and $|\psi_{-}\rangle=\exp(\mathrm{i}\alpha\hat{\Omega})|\psi_{i}\rangle$. Taking partial trace on the state $|\Psi_{f}\rangle$, the final pointer can be calculated as a mixed state:
\begin{equation}
	\hat{\rho}_{f} = \mathrm{Tr}_{\mathrm{qubit}}\left(|\Psi_{f}\rangle\langle\Psi_{f}|\right) = \cos^{2}\frac{\theta}{2}|\psi_{+}\rangle\langle\psi_{+}|+\sin^{2}\frac{\theta}{2}|\psi_{-}\rangle\langle\psi_{-}\rangle \label{eq:B2}
\end{equation}
The QFI of parameter $\theta$ on state $\hat{\rho}_{f}$ is given by:
\begin{equation}
	\mathcal{Q}(\theta) = \mathrm{Tr}\left(\hat{\rho}_{f}\hat{L}_{\theta}^{2}\right) \label{eq:B3}
\end{equation}
where $\hat{L}_{\theta}$ is the symmetric logarithmic derivative (SLD) for parameter $\theta$, and it is governed by the relation\cite{Liu_2019}:
\begin{equation}
	\hat{\rho}_{f}\hat{L}_{\theta}+\hat{L}_{\theta}\hat{\rho}_{f} = 2\partial_{\theta}\hat{\rho}_{f} \label{eq:B4}
\end{equation}

To calculate the QFI in Eq. \ref{eq:B3}, we construct a set of eigenkets $\mathcal{S}=\left\{|e_{1}\rangle,|e_{2}\rangle,\dots,|e_{\mathrm{n}}\rangle\right\}$ on the Hilbert space of state $\hat{\rho}_{f}$, where $\mathrm{n}=\mathrm{dim}(\hat{\rho}_{f})$ is the dimension of this Hilbert space, and the eigenkets $|e_{1}\rangle$, $|e_{2}\rangle$ are constructed as:
\begin{equation}
	\label{eq:B5}
	\begin{split}
		|e_{1}\rangle &= \frac{1}{\sqrt{2(1+\delta)}}\left(|\psi_{+}\rangle+|\psi_{-}\rangle\right) \\
		|e_{2}\rangle &= \frac{1}{\sqrt{2(1-\delta)}}\left(|\psi_{+}\rangle-|\psi_{-}\rangle\right)
	\end{split}
\end{equation}
where $\delta=\mathrm{Re}\langle\psi_{+}|\psi_{-}\rangle=\mathrm{Re}\langle\psi_{i}|\mathrm{e}^{\mathrm{i}2\alpha\hat{\Omega}}|\psi_{i}\rangle\approx 1-2\alpha^{2}\langle\hat{\Omega}^{2}\rangle_{i}$. Then the final pointer state can be rewritten as:
\begin{equation}
	\hat{\rho}_{f} = \frac{1+\delta}{2}|e_{1}\rangle\langle e_{1}|+\frac{1-\delta}{2}|e_{2}\rangle\langle e_{2}|+\frac{\sqrt{1-\delta^{2}}}{2}\cos\theta\left(|e_{1}\rangle\langle e_{2}|+|e_{2}\rangle\langle e_{1}|\right) \label{eq:B6}
\end{equation}
and its partial derivative of $\theta$ is
\begin{equation}
	\partial_{\theta}\hat{\rho}_{f} = -\frac{\sqrt{1-\delta^{2}}}{2}\sin\theta\left(|e_{1}\rangle\langle e_{2}|+|e_{2}\rangle\langle e_{1}|\right) \label{eq:B7}
\end{equation} 
Combing with Eq. \ref{eq:B4} and Eq. \ref{eq:B6}, we have four equations about the matrix entries of SLD $\hat{L}_{\theta}$:
\begin{equation}
	\label{eq:B8}
	\left\{
	\begin{split}
		2\langle e_{1}|\partial_{\theta}\hat{\rho}_{f}|e_{1}\rangle &= (1+\delta)\langle e_{1}|\hat{L}_{\theta}|e_{1}\rangle+\frac{\sqrt{1-\delta^{2}}}{2}\cos\theta\left(\langle e_{2}|\hat{L}_{\theta}|e_{1}\rangle+\langle e_{1}|\hat{L}_{\theta}|e_{2}\rangle\rangle\right) = 0 \\
		2\langle e_{2}|\partial_{\theta}\hat{\rho}_{f}|e_{2}\rangle &= (1-\delta)\langle e_{2}|\hat{L}_{\theta}|e_{2}\rangle+\frac{\sqrt{1-\delta^{2}}}{2}\cos\theta\left(\langle e_{1}|\hat{L}_{\theta}|e_{2}\rangle+\langle e_{2}|\hat{L}_{\theta}|e_{1}\rangle\rangle\right) = 0 \\
		2\langle e_{1}|\partial_{\theta}\hat{\rho}_{f}|e_{2}\rangle &= \langle e_{1}|\hat{L}_{\theta}|e_{2}\rangle+\frac{\sqrt{1-\delta^{2}}}{2}\cos\theta\left(\langle e_{2}|\hat{L}_{\theta}|e_{2}\rangle+\langle e_{1}|\hat{L}_{\theta}|e_{1}\rangle\rangle\right) = -\sqrt{1-\delta^{2}}\sin\theta \\
		2\langle e_{2}|\partial_{\theta}\hat{\rho}_{f}|e_{1}\rangle &= \langle e_{2}|\hat{L}_{\theta}|e_{1}\rangle+\frac{\sqrt{1-\delta^{2}}}{2}\cos\theta\left(\langle e_{1}|\hat{L}_{\theta}|e_{1}\rangle+\langle e_{2}|\hat{L}_{\theta}|e_{2}\rangle\rangle\right) = -\sqrt{1-\delta^{2}}\sin\theta
	\end{split}
	\right.
\end{equation}
from which we can calculate four matrix entries of SLD $\hat{L}_{\theta}$:
\begin{align}
	\langle e_{1}|\hat{L}_{\theta}|e_{1}\rangle &= (1-\delta)\cot\theta,\quad\langle e_{2}|\hat{L}_{\theta}|e_{2}\rangle = (1+\delta)\cot\theta \nonumber\\
	\langle e_{1}|\hat{L}_{\theta}|e_{2}\rangle &= \langle e_{2}|\hat{L}_{\theta}|e_{1}\rangle = -\sqrt{1-\delta^{2}}\csc\theta \label{eq:B9}
\end{align}

The expression of QFI in Eq. \ref{eq:B3} can be calculated by
\begin{align}
	\mathcal{Q}(\theta) &= \mathrm{Tr}\left(\hat{\rho}_{f}\hat{L}_{\theta}^{2}\right) = \mathrm{Tr}\left(\hat{L}_{\theta}\partial_{\theta}\hat{\rho}_{f}\right) = \sum_{|e_{x}\rangle\in\mathcal{S}}\langle e_{x}|\hat{L}_{\theta}\partial_{\theta}\hat{\rho}_{f}|e_{x}\rangle \nonumber\\
	&= -\frac{\sqrt{1-\delta^{2}}}{2}\sin\theta\left(\langle e_{2}|\hat{L}_{\theta}|e_{1}\rangle+\langle e_{1}|\hat{L}_{\theta}|e_{2}\rangle\right) \label{eq:B10}
\end{align}
Combining with Eq. \ref{eq:B9}, the QFI can be calculated finally as:
\begin{equation}
	\mathcal{Q}(\theta) = 1-\delta^{2} = 1-\left(\mathrm{Re}\langle\psi_{+}|\psi_{-}\rangle\right)^{2} \approx 4\alpha^{2}\langle\hat{\Omega}^{2}\rangle_{i} \label{eq:B11}
\end{equation}
where $\langle\hat{\Omega}^{2}\rangle_{i}=\langle\psi_{i}|\hat{\Omega}^{2}|\psi_{i}\rangle$ is the secondary moment of coupling operator $\hat{\Omega}$ on the initial pointer state.

\section{Generation method of HG beams and experimental results} \label{sec:A3}
Traditionally, mode cleaner cavity is necessary for generating high-order HG beams\cite{Kong:12,Chu:12}. However, mode cleaner cavity is usually difficult to setup and control in experiment. In this work, we generate the high-order HG beams by a SLM and 4-f spatial filter system\cite{Clark:16}, which is easier to implement in experiment.

In this scheme, the light beam from a $\SI{780}{\nm}$ DBR laser was expanded to a $\SI{8.6}{\mm}$-width Gaussian beam by a fiber coupler. The complex amplitude of expanded Gaussian beam is denoted as: $A_{in}(x,y)\exp\left[\mathrm{i}\phi_{in}(x,y)\right]$. Then inputting this light into a SLM, where the phase map $H(x,y)$ is displayed. The output amplitude of SLM can be denoted as:
\begin{equation}
	S(x,y) = A_{in}(x,y)\exp\left[\mathrm{i}\phi_{in}(x,y)+\mathrm{i}H(x,y)\right] \label{eq:C1}
\end{equation}
Here, we denote the relative phase as: $\phi_{r}=\phi_{out}-\phi_{in}+\phi_{g}$, where $\phi_{g}$ is the grating phase, and the relative amplitude is $A_{r}=A_{out}/A_{in}$. To filter the target light, we let
\begin{equation}
	H(x,y) = f(A_{r})\sin(\phi_{r}) \label{eq:C2}
\end{equation}
Based on the Bessel expansion formula:
\begin{equation}
	\exp\left[\mathrm{i}f(a)\sin(\phi)\right] = \sum_{-\infty}^{\infty}J_{q}\left[f(a)\right]\exp(iq\phi) \label{eq:C3}
\end{equation}
where $J_{q}[\cdot]$ is the $q$-order Bessel function. Thus, we have the amplitude of the 1st-order diffraction beam is $A_{in}\cdot J_{1}\left[f(A_{r})\right]\exp(i\phi_{out})$. The mapping function $f(\cdot)$ can be easily derived as:
\begin{equation}
	f(A_{r}) = J_{1}^{-1}(A_{r}) \label{eq:C4}
\end{equation}
where $J_{1}^{-1}(A_{r})$ is the inverse function of 1st-order Bessel function.
\begin{table}[H]
	\label{tab:C1}
	\centering
	\caption{Experimental results of generated HG beams.}
	\begin{tabular*}{\textwidth}{@{\extracolsep{\fill}}cccccccc@{\extracolsep{\fill}}}
		\toprule
		HG mode & HG00 & HG11 & HG22 & HG33 & HG44 & HG55 & HG66 \\
		\midrule
		Phase map
		& \begin{minipage}[b]{0.09\textwidth}
			\centering
			\raisebox{-.5\height}{\includegraphics[width=\textwidth]{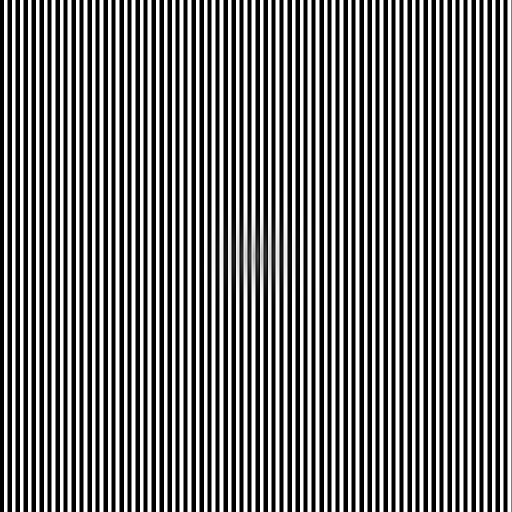}}
		\end{minipage}
		& \begin{minipage}[b]{0.09\textwidth}
			\centering
			\raisebox{-.5\height}{\includegraphics[width=\textwidth]{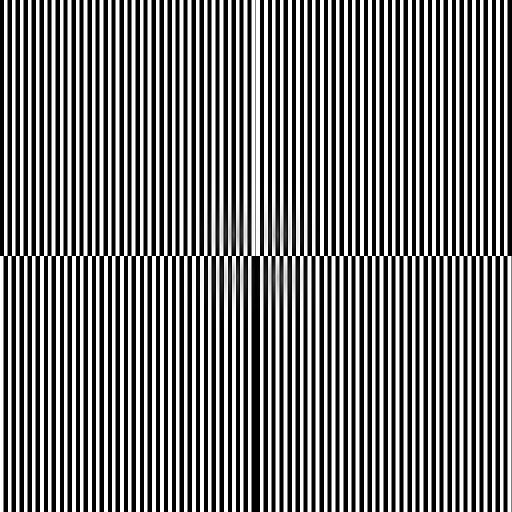}}
		\end{minipage}
		& \begin{minipage}[b]{0.09\textwidth}
			\centering
			\raisebox{-.5\height}{\includegraphics[width=\textwidth]{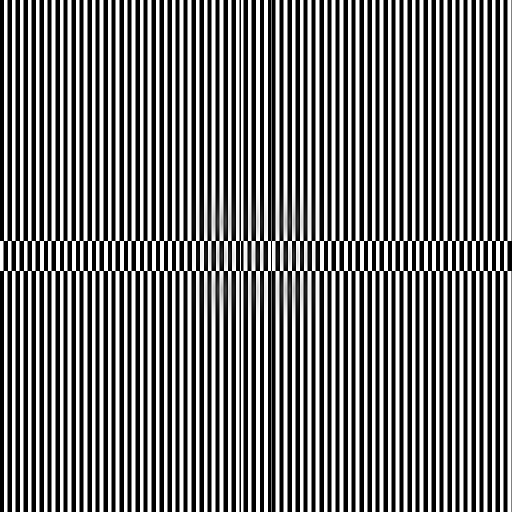}}
		\end{minipage}
		& \begin{minipage}[b]{0.09\textwidth}
			\centering
			\raisebox{-.5\height}{\includegraphics[width=\textwidth]{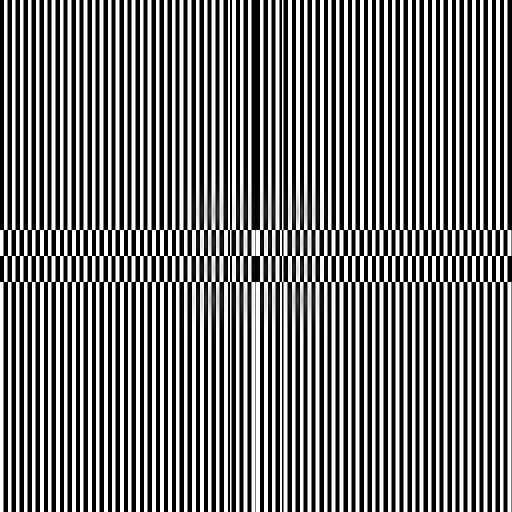}}
		\end{minipage}
		& \begin{minipage}[b]{0.09\textwidth}
			\centering
			\raisebox{-.5\height}{\includegraphics[width=\textwidth]{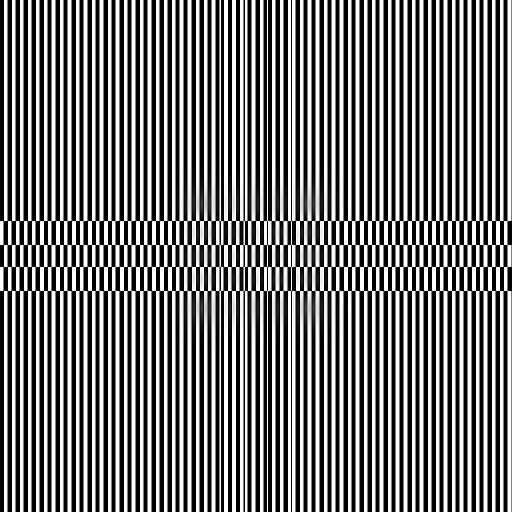}}
		\end{minipage}
		& \begin{minipage}[b]{0.09\textwidth}
			\centering
			\raisebox{-.5\height}{\includegraphics[width=\textwidth]{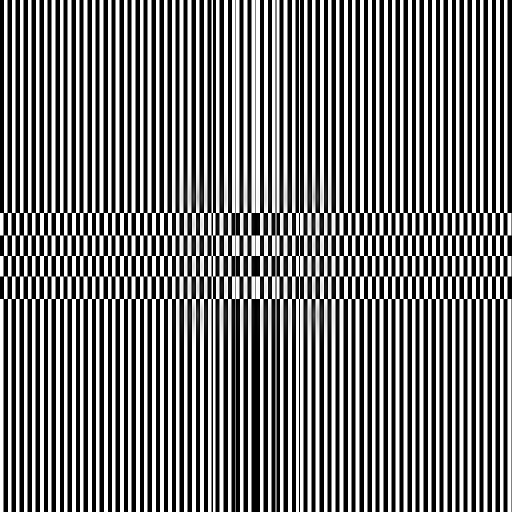}}
		\end{minipage}
		& \begin{minipage}[b]{0.09\textwidth}
			\centering
			\raisebox{-.5\height}{\includegraphics[width=\textwidth]{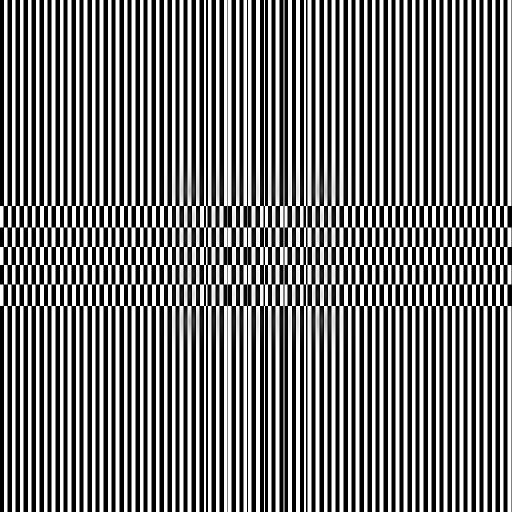}}
		\end{minipage}
		\\
		\specialrule{0em}{1pt}{1pt}
		Simulation
		& \begin{minipage}[b]{0.09\textwidth}
			\centering
			\raisebox{-.5\height}{\includegraphics[width=\textwidth]{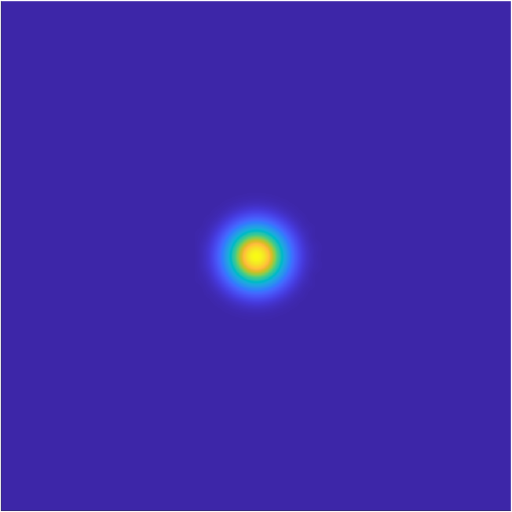}}
		\end{minipage}
		& \begin{minipage}[b]{0.09\textwidth}
			\centering
			\raisebox{-.5\height}{\includegraphics[width=\textwidth]{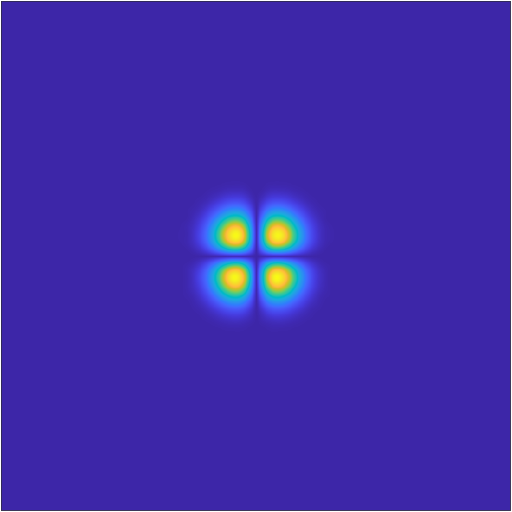}}
		\end{minipage}
		& \begin{minipage}[b]{0.09\textwidth}
			\centering
			\raisebox{-.5\height}{\includegraphics[width=\textwidth]{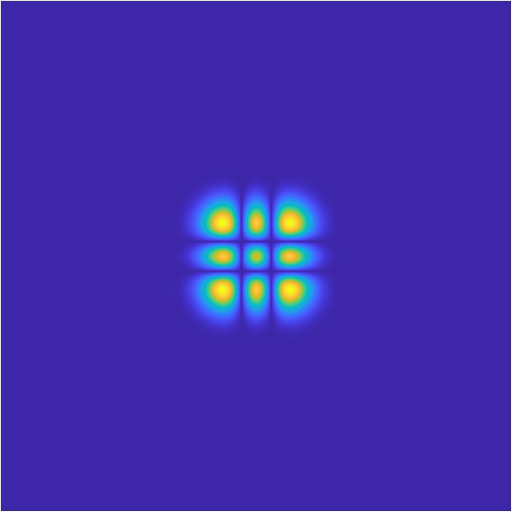}}
		\end{minipage}
		& \begin{minipage}[b]{0.09\textwidth}
			\centering
			\raisebox{-.5\height}{\includegraphics[width=\textwidth]{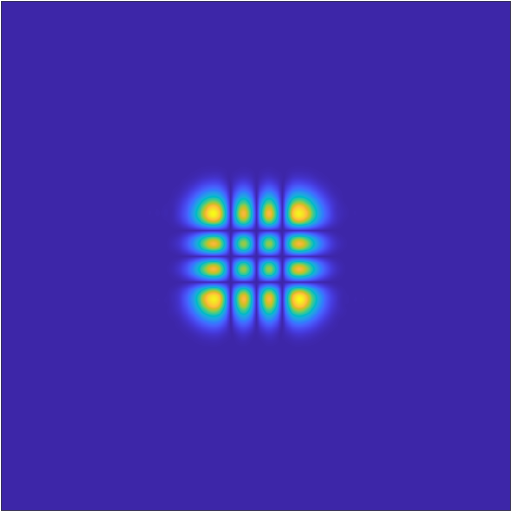}}
		\end{minipage}
		& \begin{minipage}[b]{0.09\textwidth}
			\centering
			\raisebox{-.5\height}{\includegraphics[width=\textwidth]{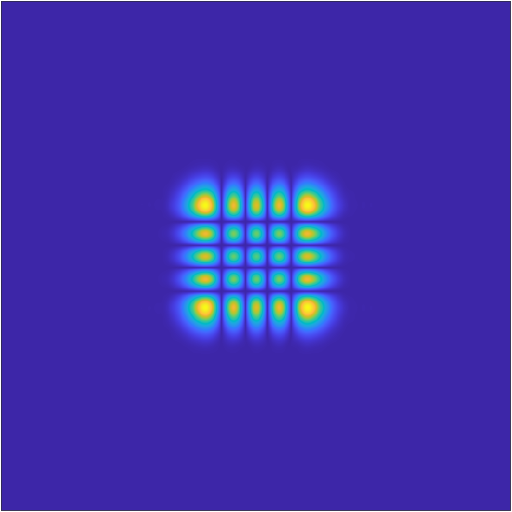}}
		\end{minipage}
		& \begin{minipage}[b]{0.09\textwidth}
			\centering
			\raisebox{-.5\height}{\includegraphics[width=\textwidth]{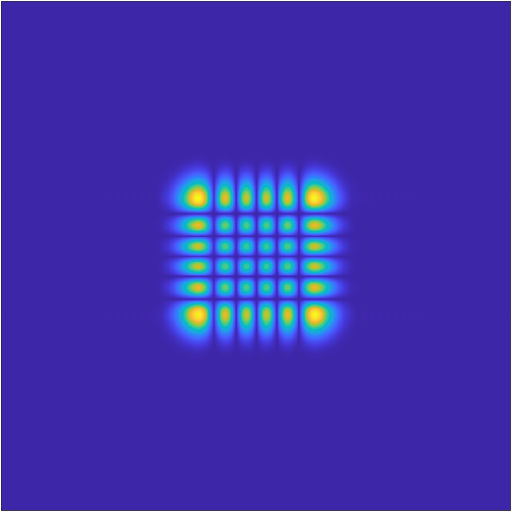}}
		\end{minipage}
		& \begin{minipage}[b]{0.09\textwidth}
			\centering
			\raisebox{-.5\height}{\includegraphics[width=\textwidth]{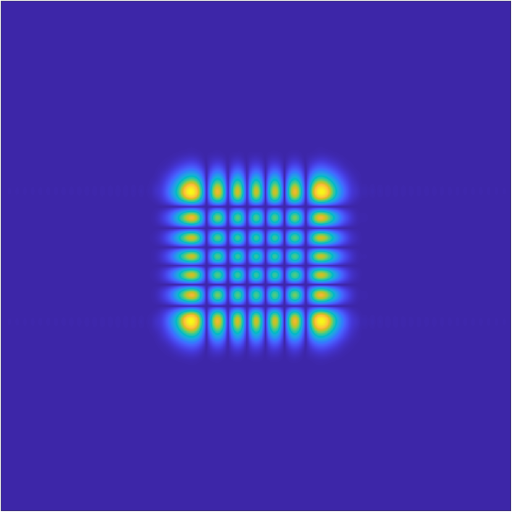}}
		\end{minipage}
		\\
		\specialrule{0em}{1pt}{1pt}
		Experiment
		& \begin{minipage}[b]{0.09\textwidth}
			\centering
			\raisebox{-.5\height}{\includegraphics[width=\textwidth]{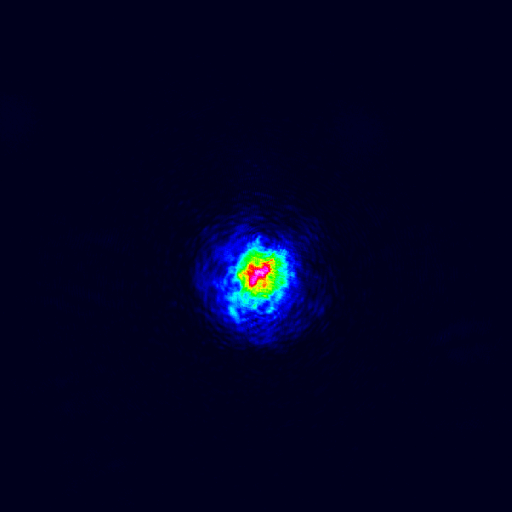}}
		\end{minipage}
		& \begin{minipage}[b]{0.09\textwidth}
			\centering
			\raisebox{-.5\height}{\includegraphics[width=\textwidth]{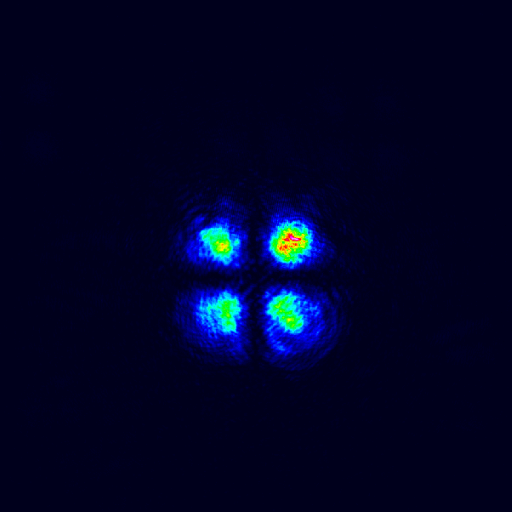}}
		\end{minipage}
		& \begin{minipage}[b]{0.09\textwidth}
			\centering
			\raisebox{-.5\height}{\includegraphics[width=\textwidth]{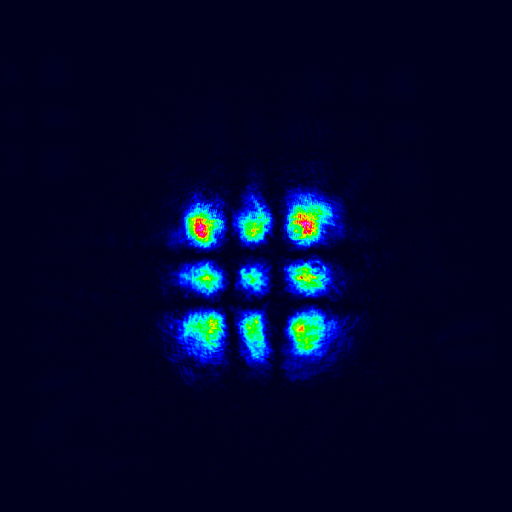}}
		\end{minipage}
		& \begin{minipage}[b]{0.09\textwidth}
			\centering
			\raisebox{-.5\height}{\includegraphics[width=\textwidth]{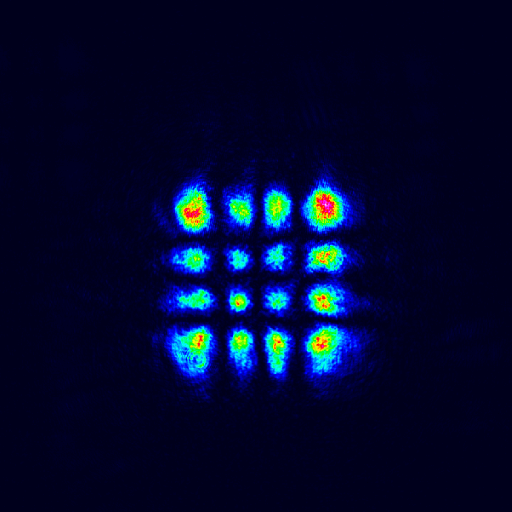}}
		\end{minipage}
		& \begin{minipage}[b]{0.09\textwidth}
			\centering
			\raisebox{-.5\height}{\includegraphics[width=\textwidth]{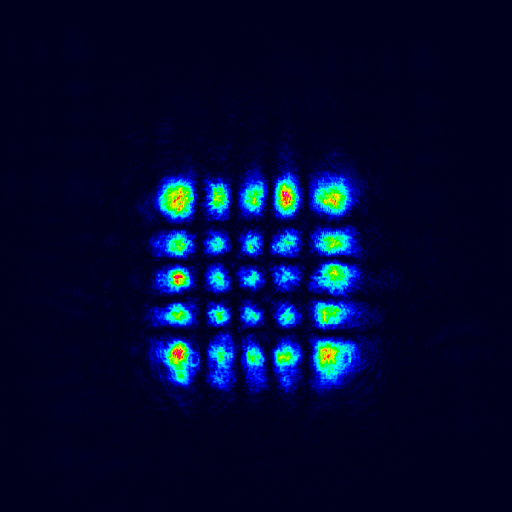}}
		\end{minipage}
		& \begin{minipage}[b]{0.09\textwidth}
			\centering
			\raisebox{-.5\height}{\includegraphics[width=\textwidth]{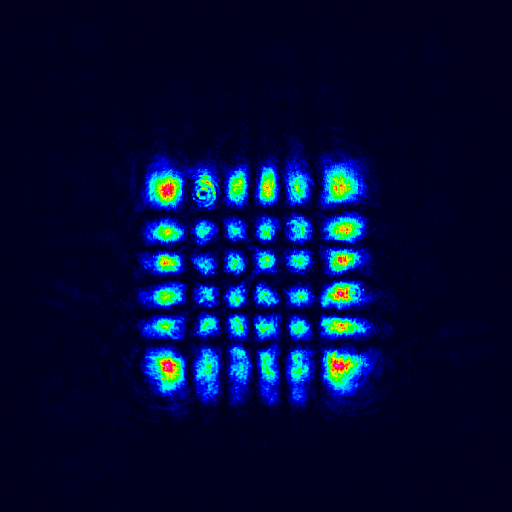}}
		\end{minipage}
		& \begin{minipage}[b]{0.09\textwidth}
			\centering
			\raisebox{-.5\height}{\includegraphics[width=\textwidth]{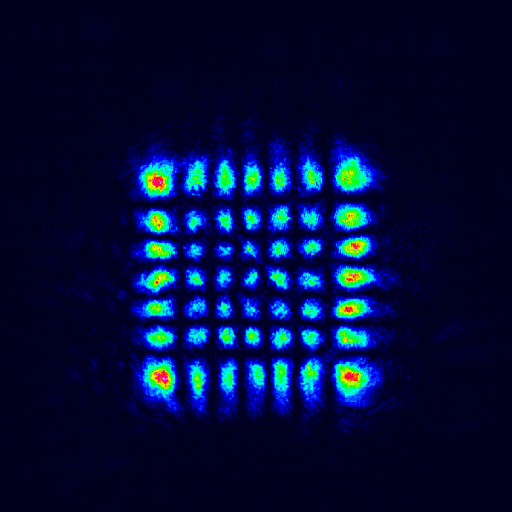}}
		\end{minipage}
		\\
		\midrule
		Purity & 97.87\% & 94.91\% & 93.76\% & 92.54\% & 90.17\% & 88.24\% & 86.75\% \\
		\bottomrule
	\end{tabular*}
\end{table}

By employing a 4-f spatial filter system with an aperture at the 1st-order diffraction point, we can generate any target beam amplitude $T(x,y)=A_{out}\exp(i\phi_{out})$ with the displayed phase map $H(x,y) = J_{1}^{-1}(A_{r})\sin(\phi_{r})$ on SLM. Here, we illustrate the experimentally generated results of HG00 to HG66 beams, and calculate the correFsponding purity in the above table.

\section{Implementation of projective measurement} \label{sec:A4}
In our experimental scheme, the rotation signal was finally detected by projective measurement\cite{PhysRevLett.112.200401}. Suppose that the input light field on SLM is $g(x,y)$, and the modulation light field on SLM is $h(x,y)$ (modulation method is same as the generation method of HG beams). The input field and the modulation field are simply combined as $g(x,y)h(x,y)$ on SLM, and a Fourier lens transfer this filed to
\begin{equation}
	f(u,v) = \mathscr{F}\left[g(x,y)h(x,y)\right] = \iint_{-\infty}^{+\infty}g(x,y)h(x,y)\exp\left[-\mathrm{i}2\pi(xu+yv)\right]\mathrm{d}x\mathrm{d}y \label{eq:D1}
\end{equation}
which is spatially filtered by a SMF coupled to an APD, the coupling efficiency into the fiber is given as:
\begin{equation}
	\eta \propto \left|\iint_{-\infty}^{+\infty}f(u,v)\exp\left(\frac{u^{2}+v^{2}}{w_{f}^{2}}\right)\mathrm{d}u\mathrm{d}v\right|^{2} \label{eq:D2}
\end{equation}
where $w_{f}$ is the field width of fiber mode. In our experiment, $w_{f}=\SI{4.6}{\um}$, which is much smaller of size scale than the features of $f(u,v)$. (The focal length of the Fourier lens is $\SI{10}{\cm}$, which transfer the waist width of $\SI{500}{\um}$ HG00 beam to nearly $\SI{50}{\um}$.) Therefore, we have $\iint f(u,v)\exp\left[(u^{2}+v^{2})/w_{f}^{2}\right]\mathrm{d}u\mathrm{d}v\approx\iint f(0,0)\exp\left[(u^{2}+v^{2})/w_{f}^{2}\right]\mathrm{d}u\mathrm{d}v$, which leads to
\begin{equation}
	\eta \propto \left|f(0,0)\right|^{2} = \left|\iint_{-\infty}^{+\infty}g(x,y)h(x,y)dxdy\right|^{2} = \left|\langle h^{*}|g\rangle\right|^{2} \label{eq:D3}
\end{equation}

From Eq. \ref{eq:D3}, we know that the detected intensity of APD directly reflected the projective probability of state $|g\rangle$ on state $|h^{*}\rangle$. Because the amplitude of $mn$-order HG beam is real, i.e. $|u_{mn}(z)\rangle=|u_{mn}^{*}(z)\rangle$. By modulating $|\psi_{\hat{L}}\rangle$ on SLM, we have the detection probability of APD is
\begin{equation}
	P = \left|\langle\psi_{\hat{L}}|\psi_{f}\rangle\right|^{2} \approx (2mn+m+n)(\cot\varepsilon)^{2}\alpha^{2} \label{eq:D4}
\end{equation}
In a practical system, the precision limit is given by is given by the classical Cram$\acute{\mathrm{e}}$r-Rao (CCR) bound\cite{PhysRevApplied.13.034023,10.1088/1361-6455/abe5c7}: $\delta\hat{\alpha}^{2}\ge 1/N\mathcal{F}(\alpha)$, where
\begin{equation}
	\mathcal{F}(\alpha) = \frac{1}{P}\left[\frac{\partial P}{\partial\alpha}\right]^{2}+\frac{1}{1-P}\left[\frac{\partial (1-P)}{\partial\alpha}\right]^{2} \approx 4(2mn+m+n)(\cot\varepsilon)^{2} \label{eq:D5}
\end{equation}
is the classical Fisher information of our projection measurement. Thus, the minimum practical detectable rotation $\alpha$ given by CCR bound is:
\begin{equation}
	\alpha_{\min}^{\mathrm{CCR}} = \frac{1}{\sqrt{2mn+m+n}} \frac{1}{2|\cot\varepsilon|\sqrt{N}} \label{eq:D6}
\end{equation}
which determines that $\alpha_{\min}^{\mathrm{CCR}}=\alpha_{\min}^{\mathrm{QCR}}$. In another word, the significant enhancement on measurement precision can be achieved in a practical optical system without involving any quantum resources.

\begin{backmatter}
\bmsection{Funding}
National Natural Science Foundation of China Grants No.62071298, No. 61671287, No.61631014, and No.61901258.

\bmsection{Acknowledgments}
We thank Lijian Zhang and Kui Liu for the helpful discussions. This work was supported by the National Natural Science	Foundation of China (Grants No.62071298, No. 61671287, No.61631014, and No.61901258) and the fund of the State Key Laboratory of Advanced Optical Communication Systems and Networks.

\bmsection{Disclosures} The authors declare no conflicts of interest.

\bmsection{Data availability} Data underlying the results presented in this paper are not publicly available at this time but may be obtained from the authors upon reasonable request.

\end{backmatter}

%%%%%%%%%%%%%%%%%%%%%%% References %%%%%%%%%%%%%%%%%%%%%%%%%

%%%%%%%%%% If using BibTeX:
\bibliography{bibliography}

%%%%%%%%%% If preparing manually:
% \newcommand{\enquote}[1]{``#1''}

% \bibitem{Zhang:14}
% Y.~Zhang, S.~Qiao, L.~Sun, Q.~W. Shi, W.~Huang, L.~Li, and Z.~Yang,
%   \enquote{Photoinduced active terahertz metamaterials with nanostructured
%   vanadium dioxide film deposited by sol-gel method,}
%   {\protect\JournalTitle{Optics Express}} \textbf{22}, 11070--11078 (2014).

% {Optical Society}, \enquote{{OSA Publishing},}
%   \url{http://www.osapublishing.org}.

% \bibitem{FORSTER2007}
% P.~Forster, V.~Ramaswamy, P.~Artaxo, T.~Bernsten, R.~Betts, D.~Fahey,
%   J.~Haywood, J.~Lean, D.~Lowe, G.~Myhre, J.~Nganga, R.~Prinn, G.~Raga,
%   M.~Schulz, and R.~V. Dorland, \enquote{Changes in atmospheric consituents and
%   in radiative forcing,} in \enquote{Climate Change 2007: The Physical Science
%   Basis. Contribution of Working Group 1 to the Fourth assesment report of
%   Intergovernmental Panel on Climate Change,}  S.~Solomon, D.~Qin, M.~Manning,
%   Z.~Chen, M.~Marquis, K.~B. Averyt, M.~Tignor, and H.~L. Miler, eds.
%   (Cambridge University Press, 2007).

% \end{thebibliography}

\end{document}